\documentclass[11pt,a4paper]{article}
\usepackage{jheppub}
\usepackage{natbib}
\usepackage{graphicx}
\usepackage{cases}
\usepackage[dvipsnames]{xcolor}

\usepackage{mathtools}
\usepackage{hyperref}
\usepackage{float}
\usepackage{comment}
\usepackage{multirow}
\usepackage[overload]{empheq}
\usepackage[normalem]{ulem}
\usepackage{orcidlink}
\usepackage{rorlink}
\usepackage{bm}
\usepackage{tikz}
\usepackage{slashed}
\usepackage[compat=1.1.0]{tikz-feynman}

\definecolor{Blu}{rgb}{0.,0.,1.}

\author[a]{Marco Cirelli$^{\orcidlink{0000-0002-4264-6323}}$} 
\author[a]{Arpan Kar$^{ \orcidlink{0000-0002-2993-3336}}$} 
\author[b]{Antoine Letessier-Selvon$^{\orcidlink{0000-0002-2044-3103}}$} 
\author[b]{Harrys Lumengo-Kidimbu$^{\orcidlink{0009-0008-1680-3285}}$}

\affiliation[a]{Laboratoire de Physique Théorique et Hautes Énergies (LPTHE)$^{\ \rorlink{https://ror.org/02mph9k76}}$, \\ 
CNRS $\&$ Sorbonne Université, \\ 4 Place Jussieu, Paris, France}
\affiliation[b]{Laboratoire de Physique Nucléaire et de Hautes Énergies (LPNHE)$^{\ \rorlink{https://ror.org/01hg8p552}}$, \\ 
Sorbonne Université, Université de Paris, CNRS-IN2P3, \\ 4 Place Jussieu, Paris, France}

\emailAdd{marco.cirelli@gmail.com}
\emailAdd{arpankarphys@gmail.com}
\emailAdd{antoine.letessier-selvon@in2p3.fr}
\emailAdd{hlumengo@lpnhe.in2p3.fr}

\title{Direct and Indirect searches for DM-electron interactions in sub-GeV DM models}

\abstract{We analyze the complementarity between direct detection (DD) and indirect detection (ID) 
searches of sub-GeV dark matter (DM) in the context of two realistic models: the vector-portal (dark photon) and scalar-portal (higgs-portal) models. 
For DD we focus on the latest constraints on the DM-electron scattering from the present {\sc Damic-M} experiment  
as well as its future projection. 
For ID we consider several existing 
X-rays/gamma-rays and cosmic-ray ($e^\pm$) observations as well as the 
upcoming MeV telescope {\sc Cosi} to obtain the corresponding bounds and projections 
on the DM annihilation rate, which are then translated 
in terms of the DM-electron scattering cross-section and compared with the DD bounds/projections. 
We properly take into account astrophysical and galactic propagation uncertainties. 
We find that the complementarity between {\sc Damic-M} and ID works best for 
the vector-portal model, especially for the case where the vector mediator 
is heavier than the DM particle: in this case, the present {\sc Damic-M} bound becomes comparable to 
(or better than, for some astrophysical configurations) the combined ID limit 
for $3\,{\rm MeV} \lesssim m_{\rm DM} \lesssim 20\,{\rm MeV}$. 
For other cases and mass ranges, the ID is in general more constraining and can also, for some mass ranges, 
constrain the parameter space below the neutrino floor, which is inaccessible to DD experiments.}

\begin{document}

\maketitle

\section{Introduction}
\label{sec:introduction}

Identifying the nature of Dark Matter (DM) remains one of the most pressing challenges in particle physics and cosmology today (see e.g.~\cite{Cirelli:2024ssz} for a recent overview). The main difficulty is that the astrophysical and cosmological phenomena that inform us of the very existence of DM, through its gravitational effects, provide essentially no clues as for its mass and kinds of interactions (other than gravity). The searches therefore span huge orders of magnitude in mass and interaction strength.

\medskip

In this broad context, the case of sub-GeV DM, intended as a particle with a mass $m_{\rm DM}$ in the range of roughly 1 MeV to a few GeVs, has emerged as one of the most active frontiers in DM phenomenology. This is due, on one side, to the fact that the more classical weak-scale DM has so far failed to show up in past and current experiments, despite the significant improvements in sensitivity achieved in the recent decades. On the other side, to the fact that several theoretically well-motivated models encompassing sub-GeV DM have recently been proposed in the literature (see e.g.~\cite{Boehm:2002yz, Boehm:2003hm, Boehm:2003bt, Fayet:2007ua, Feng:2008ya, Hochberg:2014dra, Hochberg:2014kqa, Bertuzzo:2017lwt, Knapen:2017xzo, Darme:2017glc} for a selection). The search for sub-GeV DM is thus a still rather uncharted territory, where significant progress is possible. 

\medskip

When searching for particle DM, one is confronted to the usual standard strategies: direct detection (DD), i.e.~looking for DM-induced scatterings of nuclei or electrons in controlled underground experiments; indirect detection (ID), i.e.~seeking excesses in cosmic rays, photons and neutrinos induced by DM annihilations or decays in astrophysical environments; and possibly accelerator-based, laboratory searches. 
In a full model-independent approach, it is not possible to map the constraints obtained via one strategy onto another one. The relevant parameters and the kinematical conditions proper to the signals in each strategy are different and, in general, not necessarily related. 
In specific models, however, relating the different approaches is not only possible, but often very beneficial, since different areas of the parameter spaces can be covered by different probes, thus fully exploiting the complementarity of the searches. 
This case is what we aim at, in the present work. 

On the ID side, we make use of the results of ref.~\cite{Cirelli:2025rky}, which focused on some representative realistic sub-GeV DM models and performed a detailed study of the ID constraints based on the existing photon and cosmic-ray observations by {\sc Comptel, Integral, Fermi-Lat}, {\sc Voyager-1} and {\sc AMS-02}. It also evaluated the future prospects from upcoming MeV photon telescopes like {\sc Cosi}. 

On the DD side, we focus on the {\sc Damic-M } experiment, which uses skipper charge-coupled devices (CCDs) to measure possible DM-electron scatterings. Using a prototype detector in the Low Background Chamber (LBC) of the Laboratoire Souterrain de Modane (LSM), the experiment  established the most stringent limits on the search for sub-GeV DM in ref.~\cite{DAMIC-M:2025luv}. 
The final {\sc Damic-M} detector aims at improving these results by up to 3 orders of magnitude \cite{DAMIC-M:2022aks}. 

The purpose of the present work is to combine the two studies and derive comprehensive and updated constraints on the representative realistic sub-GeV DM models of \cite{Cirelli:2025rky}. 

\bigskip

The rest of this paper is organized as follows. In section \ref{sec:formalism} we recall the formalism for direct (subsection \ref{sec:formalism_DD}) and indirect (subsection \ref{sec:formalism_ID}) signals of sub-GeV DM. In sections \ref{sec:model_vector} and \ref{sec:model_scalar} we introduce the models under consideration, and provide the results from DD and ID, distinguishing a few sub-cases. 
Finally, in section \ref{sec:conclusions} we draw our conclusions.

\section{Direct and Indirect detection signals induced by sub-GeV DM}
\label{sec:formalism}

\subsection{Direct Detection}
\label{sec:formalism_DD}

For the DD of sub-GeV DM we focus on the present {\sc Damic-M} 
experiment~\cite{DAMIC-M:2025luv} 
(as well as its future projection~\cite{DAMIC-M:2022aks}) 
which is based on scattering between DM and electron using skipper charge-coupled devices (CCDs). The considered results are from the prototype of the future {\sc Damic-M} detector, the Low Background Chamber (LBC) \cite{Arnquist_2024}, located at the Modane Underground Laboratory (LSM). 

The rate of DM-electron (DM-$e$) scattering in a detector like {\sc Damic-M} can be expressed as \cite{Essig:2015cda}: 
\begin{equation}
\frac{dR}{dE_e} = \alpha N_{\rm cell} \, \bar{\sigma}_e  \frac{m_e^2}{\mu_{{\rm DM}e}^2} \left (\frac{\rho^{\rm DM}_\odot}{m_{\rm DM}}  \right) \int \frac{dq}{q^2} \left[ \int \frac{f({\bf v}_{\rm DM})}{v_{\rm DM}} d^3v_{\rm DM} \right] |F_{\rm DM}(q)|^2 |F_{c}(q, E_e)|^2
\end{equation}
Here $\bar{\sigma}_e$ is a reference scattering cross-section 
(evaluated at a fixed momentum transfer $q_{\rm ref}=\alpha\,m_e$, with $\alpha$ the usual electromagnetic fine structure constant and $m_e$ the mass of the electron), whose expression will be detailed in sections \ref{sec:model_vector} and \ref{sec:model_scalar} as it depends on the chosen particle physics model. $N_{\rm cell}=M_{\rm crystal}/M_{\rm cell}$ is the number of unit crystal cells, each of mass $M_{\rm cell}$ (52.33 GeV in Silicium), in the crystal target of mass $M_{\rm crystal}$. $\rho^{\rm DM}_\odot$ is the local DM density and $v_{\rm DM}$ is the incoming DM velocity, distributed according to the velocity distribution $f({\bf v}_{\rm DM})$ in the galactic halo, for which we will assume a Maxwell-Boltzmann function with parameters detailed in \cite{Baxter_2021}.
$F_{c}(q, E_e)$ is the product of the crystal form factor (a solid-state physics term measuring the number of electrons ionized by a given DM particle with momentum $q$ and energy $E_e$) and the screening factor (measuring in-medium effects) (see \cite{PhysRevD.109.115008}). The expression for the DM form factor $F_{\rm DM}(q)$ varies depending on the mass $m_{\rm med}$ of the hidden sector mediator \cite{Emken:2019tni, Stratman:2026qnh, Barman:2024lxy}:
\begin{equation}
F_{\rm DM} (q) = \frac{q^2_{\rm ref} + m^2_{\rm med}}{q^2 + m^2_{\rm med}} \, .
\end{equation}

\subsection{Indirect Detection} 
\label{sec:formalism_ID}

In this section we provide a brief highlight of different types of 
ID signals produced in the present universe 
by our candidate sub-GeV DM models, as well as the corresponding 
astrophysical observations used to constrain such signals. 
We follow closely the previous work \cite{Cirelli:2025rky}, to which readers are referred for 
all the technical details. 

\subsubsection{Photon signals from DM annihilation in the inner Galaxy}

\begin{itemize}
\item \underline{\bf Prompt $\gamma$-ray emission:} 
The flux produced by prompt emission from the pair-annihilation of galactic DM particles 
(averaged over the observation region $\Delta \Omega$) is given by \cite{Cirelli:2025rky}: 
\begin{equation}
\frac{d\Phi_{\rm prompt}}{dE_\gamma} = \frac{\langle \sigma v \rangle}{8 \pi \, f_{\rm DM} \, m^2_{\rm DM}} \, \left.\frac{dN_\gamma}{dE_\gamma}\right\vert_{\rm tot} \, 
\frac{J_{\Delta \Omega}}{\Delta \Omega} \, ,
\label{eq:prompt_flux}
\end{equation}
with $f_{\rm DM} =2$ (since in this work we will consider a Dirac fermion DM). 
Here $\langle \sigma v \rangle$ is the total velocity-averaged annihilation cross-section of DM particles and $\left.\frac{dN_\gamma}{dE_\gamma}\right\vert_{\rm tot}$ is the total photon energy spectrum produced per annihilation of DM particles of mass $m_{\rm DM}$ under the given DM model\footnote{The energy spectra of photon and $e^\pm$ 
produced per DM annihilation in the given model 
are obtained from the publicly-available 
package \texttt{HAZMA}~\cite{Coogan:2019qpu, Coogan:2022cdd}}. 
The annihilation $J$-factor for the observational region of interest (ROI) $\Delta \Omega$ is defined as usual as 
\begin{equation}
J_{\Delta \Omega} = \int_{\Delta \Omega} d\Omega \, \int_{\rm l.o.s.} ds \,\, \rho^2_{\rm DM} (r(s,\theta)) \, \, ,
\end{equation}
where $\rho_{\rm DM}(r)$ is the density distribution of DM in the Galaxy and 
$s$ is the line-of-sight (l.o.s.) coordinate at an angular distance $\theta$ 
from the galactic center (GC). 

\item \underline{\bf Secondary photon signals:} 
We also consider the secondary photon signals produced from galactic DM annihilations. 
Such secondary signals arise due to Inverse Compton Scatterings (ICS), 
bremsstrahlung (brem) and in-flight annihilation (IfA) 
of DM induced positrons and/or electrons, and 
can be expressed in general as: 
\begin{equation}
\frac{d\Phi_{\rm sec}}{dE_\gamma} = 
\frac{1}{\Delta \Omega} \int_{\Delta \Omega} d\Omega \, 
\left[\frac{1}{E_\gamma} \int_{\rm l.o.s.} ds \,\, \frac{j_{\rm sec} (E_\gamma, \Vec{x}(s,b,l))}{4\pi} \right] \, .
\label{eq:secondary_flux}
\end{equation}
Here $j_{\rm sec}$ is the emissivity (at a galactic position $\Vec{x}$ determined by the 
l.o.s.~coordinate $s$ and latitude and longitude $(b, l)$) corresponding to one of 
the secondary processes and is related to 
the power $\mathcal{P}_{\rm sec}$ associated to that process as:
\begin{eqnarray}
j_{\rm sec} (E_\gamma, \Vec{x}(s,b,l)) &=& \eta \, \int^{m_{\rm DM}}_{m_e} dE_e \, 
\mathcal{P}_{\rm sec} (E_\gamma, E_e, \Vec{x}) 
\,\, \frac{dn_{e}}{dE_e}(E_e,\Vec{x}) \, , 
\label{eq:j_sec}
\end{eqnarray}
with $\eta = 2$ (for ICS and brem) and 1 (for IfA). 
The $e^\pm$ distribution $\frac{dn_{e}}{dE_e}$ in the Galaxy is generated from the 
DM annihilation induced source function:
\begin{equation}
Q_e (E^S_e, r) = \frac{\langle \sigma v \rangle}{2 \, f_{\rm DM} \, m^2_{\rm DM}} \, 
\left.\frac{dN_e}{dE^S_e}\right\vert_{\rm tot} \, \rho^2_{\rm DM}(r) \, ,
\end{equation}
where 
$\left.\frac{dN_e}{dE^S_e}\right\vert_{\rm tot}$ 
is the energy spectrum of $e^\pm$ sourced by the annihilation 
of DM of mass $m_{\rm DM}$ in a given model. 
We follow \cite{Cirelli:2025rky} for the details of the computation of these secondary signals. 
We obtain $\frac{dn_{e}}{dE_e}(E_e,\Vec{x})$ using a semi-analytic approach 
and assume a mask (applied to both secondary and prompt photon signals) 
of $|b| \le 4^{\circ}$, as described in \cite{Cirelli:2025rky}. 
\end{itemize}

\noindent For the DM distribution in the Galaxy, we consider a standard NFW profile 
\begin{equation}
\rho^{\rm NFW}_{\rm DM}(r) = \frac{\rho_s}{\left(\frac{r}{r_s}\right) 
\left(1 + \frac{r}{r_s}\right)^2} \, ,
\label{eq:rho_NFW}
\end{equation} 
with the same values of the profile parameters ($\rho_s$ and $r_s$) 
used in \cite{Cirelli:2025rky}. These parameters lead to a local DM density 
$\rho^{\rm DM}_\odot \simeq 0.4$ $\rm GeV\,cm^{-3}$. 
We consider this as our benchmark DM profile. 
At the same time we also present our results considering a cored-isothermal profile
\begin{equation}
\rho^{\rm Iso}_{\rm DM}(r) = \frac{{\rho_s}}{\left(1 + \left(\frac{r}{r_s}\right)^2\right)} \, ,
\label{eq:rho_Iso}
\end{equation}
with the same values of the profile parameters considered in \cite{Cirelli:2025rky} 
(see their Appendix B). In the case of the NFW profile, we also consider 
two extreme limits of the local DM density $\rho^{\rm DM}_\odot$, 
i.e.~0.2 and 0.7 $\rm GeV\,cm^{-3}$ (see ~\cite{Cirelli:2024ssz}), 
and adjust the parameters $\rho_s$ and $r_s$ accordingly so that total mass 
contained in the Galactic halo (within 200 kpc) remains the same, i.e., 
$M_{\rm DM} (r\le200\,{\rm kpc}) \simeq 0.75\times10^{12}\,M_\odot$, as 
obtained for the benchmark NFW profile adopted in \cite{Cirelli:2025rky}. 

\subsubsection{DM signals in charged cosmic-rays}
\label{sec:DM_CR_signal}
We also consider the cosmic-ray (CR) $e^\pm$ signals arising from the 
annihilation of sub-GeV DM particles in the Milky Way. 
For the propagation in the galactic environment of the DM induced $e^\pm$ we assume the two propagation setups 
(both from \cite{Orlando:2017mvd}), named as `prop.~a' 
(which includes a small amount of reacceleration) and `prop.~c' (which includes a relatively more significant reacceleration), as considered in \cite{Cirelli:2025rky}. The propagation is dealt with 
numerically using the package \texttt{DRAGON}~\cite{Evoli:2016xgn, Evoli:2017vim}. 
For details see \cite{Cirelli:2025rky}. 

\subsubsection{Present and upcoming astrophysical observations}
\label{sec:astro_obs}

\begin{itemize}
\item \underline{\bf Existing $X$-ray/$\gamma$-ray and CR observations:} 
Following \cite{Cirelli:2025rky}, we derive constraints 
on the sub-GeV DM models by comparing the DM annihilation induced photon 
and the CR signals with the existing $X$-ray/$\gamma$-ray data from 
{\sc Integral}, {\sc Comptel}, {\sc Egret} and {\sc Fermi-Lat} 
\cite{2011ApJ...739...29B, Essig:2013goa, KappadathCOMPTEL, Strong:2004de, 2017ApJ...840...43A}, 
and the cosmic-ray $e^\pm$ data from {\sc Voyager-1}~\cite{Boudaud:2016mos, doi:10.1126/science.1236408, 2016ApJ...831...18C}, respectively. 
In addition, here we also consider the $e^\pm$ data from {\sc Ams-02}, 
extracted from the `Cosmic-Ray Data Base ({\sc Crdb})'~\cite{AMS02} 
(for the period 2011 - 2018) \cite{2021PhR...894....1A}, and 
assume a solar modulation of $\phi_F \simeq 600$ MV in the $e^\pm$ CR flux. 

For the comparison of the DM annihilation signal with the 
corresponding existing astrophysical observations we took the 
{\bf `Conservative approach'} discussed in Sec. 5.2 of \cite{Cirelli:2025rky} and 
present our main results based on this. It basically consists in assuming nothing on the astrophysical backgrounds
and compare the DM-induced signals directly to the observed data.
In parallel, for illustration, we also present the DM indirect detection constraints 
(related to the $X$-ray/$\gamma$-ray observations) obtained based on the {\bf `Optimistic approach'} 
as discussed in \cite{Cirelli:2025rky}. For this latter approach, we consider 
the fiducial astrophysical $X$-ray/$\gamma$-ray backgrounds 
for {\sc Integral}, {\sc Comptel}, {\sc Egret} and {\sc Fermi-Lat} from 
\cite{2011ApJ...739...29B}, \cite{Bartels:2017dpb}, \cite{Strong:2004de} 
and \cite{Orlando:2017mvd}, respectively. 

In the next sections, for the $X$-ray/$\gamma$-ray bound we consider the 
combined upper-limit from {\sc Integral}, {\sc Comptel}, {\sc Egret} and {\sc Fermi-Lat} 
and for the CR limit we consider the combined limit from 
{\sc Voyager-1} and {\sc Ams-02}. 

\item \underline{\bf Upcoming MeV telescope ({\sc Cosi}):} 
We also consider the projections of the upcoming space-based MeV $\gamma$-ray telescope 
{\sc Cosi} \cite{Tomsick:2019wvo, Beechert:2022phz}
in probing the parameter space of the sub-GeV DM models. {\sc Cosi}, which has already been 
selected to fly, is a space-based wide field-of-view telescope 
with a high-resolution spectroscopy and an operational energy range 0.2 -- 5 MeV. 
We estimate its projected sensitivity on our sub-GeV DM models 
following the methodology of \cite{Cirelli:2025rky}, which is 
similar to the one adopted in \cite{Caputo:2022dkz}. 
We consider a disk of radius 
$10^\circ$ around the GC and obtain the $3\sigma$ projected sensitivity 
for 2 yrs of mission time, corresponding to 
approximately 1 yr of observation time \cite{Negro:2021urm}. 
Other future MeV telescopes such as 
{\sc Amego}~\cite{AMEGO:2019gny, Caputo:2022xpx}, 
{\sc e-Astrogam}~\cite{e-ASTROGAM:2017pxr}, {\sc Gecco}~\cite{Orlando:2021get}, 
{\sc Grams}~\cite{Aramaki:2021o5}, {\sc Mast}~\cite{Dzhatdoev:2019kay} 
and {\sc MeVCube}~\cite{Saha:2025wgg} 
are expected to improve such sensitivities even further \cite{Cirelli:2025qxx}. 
\end{itemize}

\section{Kinetically mixed dark photon portal model} 
\label{sec:model_vector}

We consider sub-GeV DM particles with a mass $m_{\rm DM}$ in the range 1 MeV -- 1 GeV 
in the so-called `dark photon portal model'~\footnote{Note that, apart from the 
dark photon portal model, there are also other popular vector portal models, 
such as the B-L model (see \cite{Coogan:2022cdd, Dutra:2025cwn}). We provide 
in appendix~\ref{sec:other_vec_portal} a brief discussion related 
to its phenomenology.}. Here the dark sector consists of a fermion DM which is neutral under the SM gauge group and interacts with the SM through a new $U(1)_D$ gauge boson, 
the dark photon, which kinematically mixes with the SM photon. 
The Lagrangian corresponding to the considered dark sector is \cite{Emken:2019tni, LDMX:2018cma}: 
\begin{equation}
\mathcal{L_{D}}^V \, = \, \bar{\chi} (i \gamma^\mu D_\mu - m_{\rm DM}) \chi \, -\, \frac{1}{4} V_{\mu\nu}V^{\mu\nu} \, + \, m^2_V V_{\mu} V^{\mu} \, + \, \frac{\epsilon}{2} 
F_{\mu\nu}V^{\mu\nu}
\end{equation}
with $D_\mu = \partial_\mu - i g_D V_\mu$ and the coupling of the dark photon with DM 
$g_D = \sqrt{4 \pi \alpha_D}$. Here $\chi$ represents the DM particle (a Dirac fermion) 
which is symmetric in terms of particle-antiparticle abundances, $V$ denotes the dark photon (with a mass $m_V$) and $V^{\mu\nu}$ its field  tensor. 
The parameter $\epsilon$ represents the kinetic mixing between the dark 
and visible photons.

In this model, the reference DM-$e$ scattering cross-section 
$\bar{\sigma}_e$ and the DM form factor $F_{\rm DM}(q)$ are expressed as 
\cite{Emken:2019tni, Cheek:2025nul}:
\begin{equation}
\bar{\sigma}_e = \frac{16 \pi \, \alpha \, \alpha_D \, \epsilon^2 \, \mu^2_{{\rm DM} e}}
{\left(\alpha ^2 m_e ^2+ m^2_{V}\right)^2} 
\label{eq:sigma_e}
\end{equation}
and 
\[
F_{\rm DM}(q)=
\begin{cases}
\displaystyle 1 & \text{if } m_{V} \gg \alpha m_e \\
\displaystyle \left( \frac{\alpha \, m_e}{q} \right)^2  & \text{if } m_{V} \ll \alpha m_e 
\end{cases}
\] 
with $\mu_{{\rm DM} e}$ the reduced mass of the DM-$e$ system.

\subsection{`Heavy' dark photon: $m_{V}> m_{\rm DM}$} 
\label{sec:heavy_DP}

\begin{figure*}[!t]
\hspace{-16mm} \includegraphics[width=0.6\textwidth]{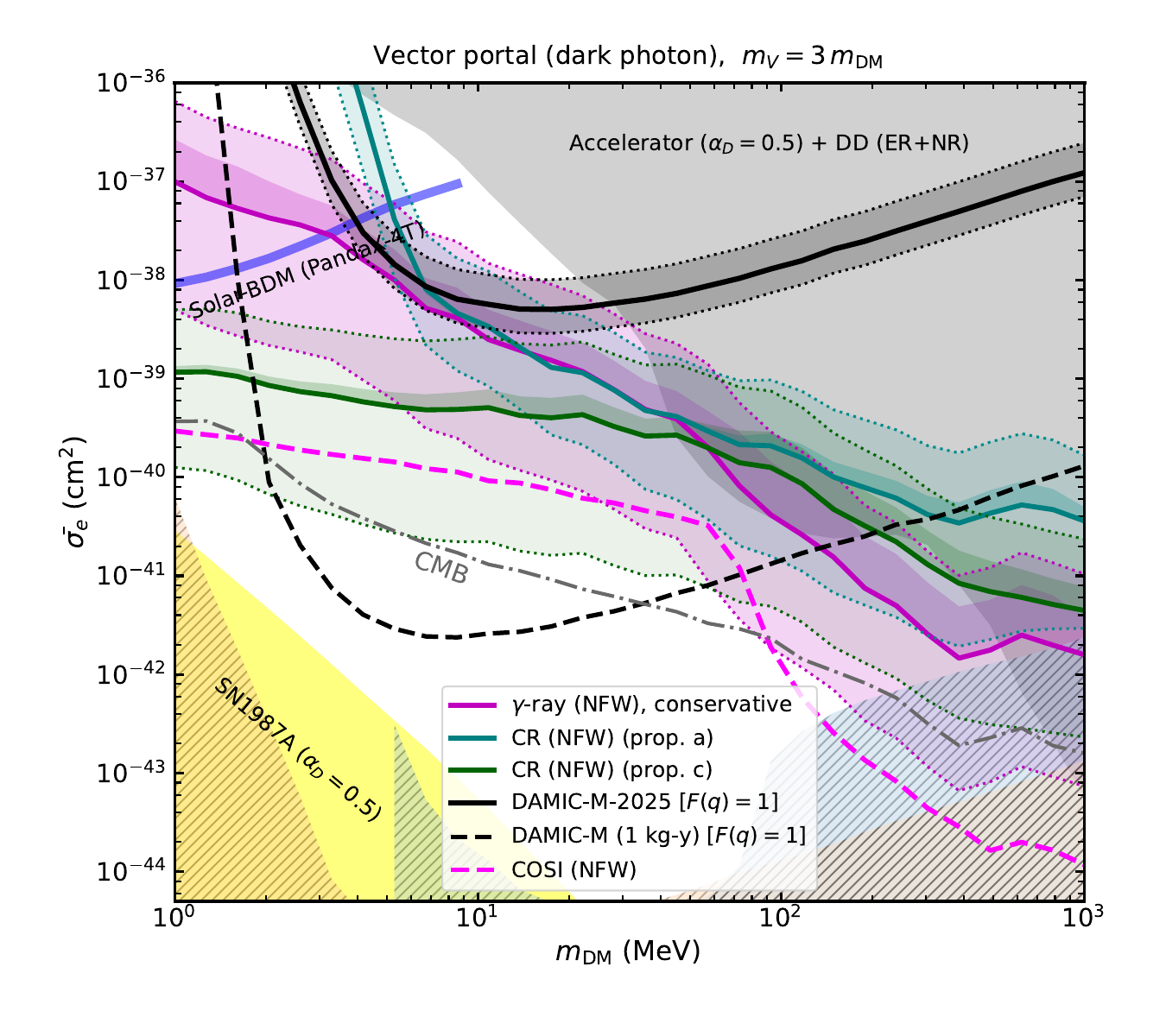}\hspace{-4mm}
\includegraphics[width=0.6\textwidth]{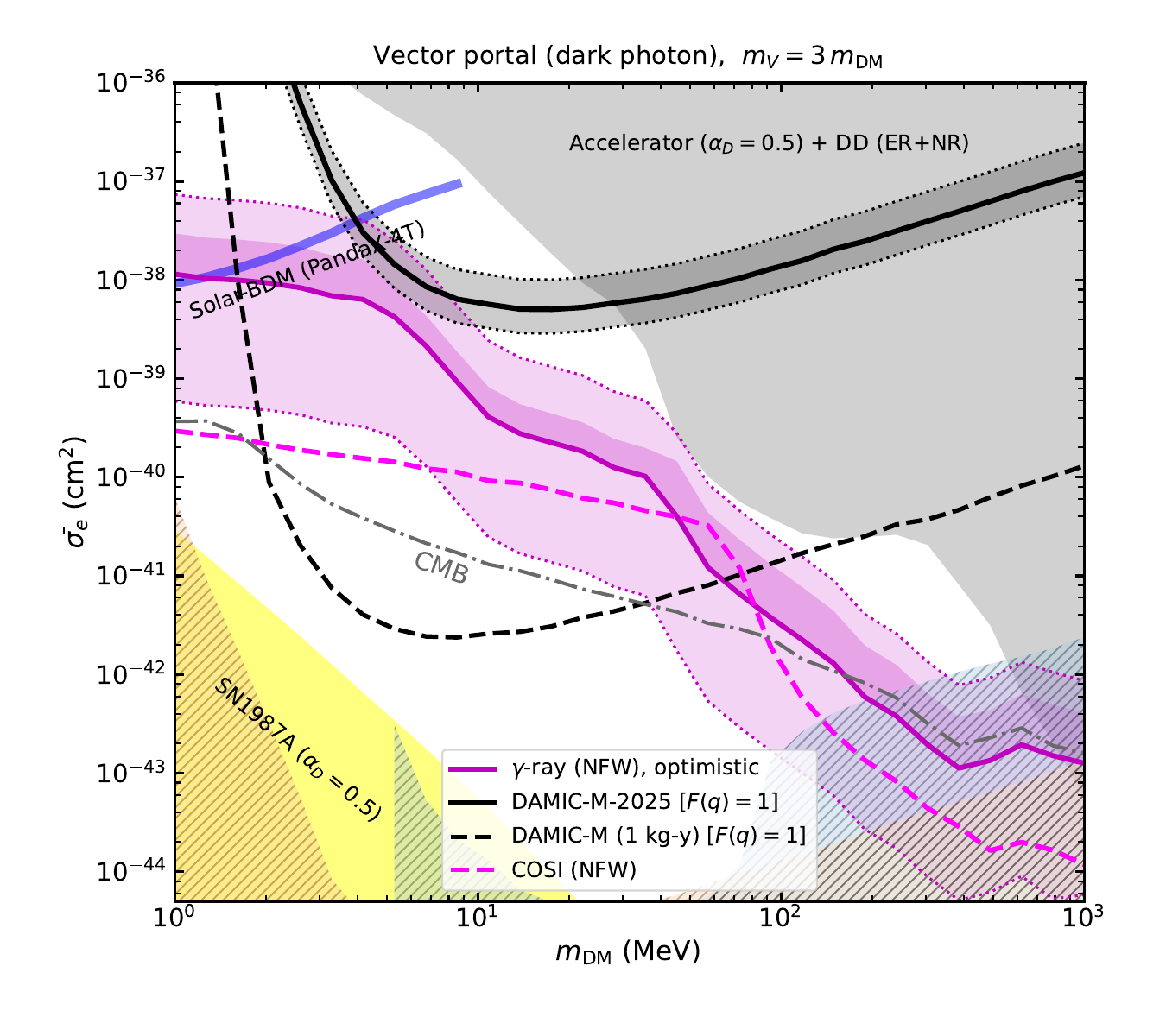}\\
\vspace{-5mm}
\caption{\em Comparisons of ID and DD constraints for the the {\bfseries vector portal model}, with a {\bfseries heavy dark photon}. 
{\bfseries Left panel}: The ID exclusions corresponding to existing $X$-ray/$\gamma$-ray and cosmic-ray 
observations (under the `conservative approach' specified in Sec.~\ref{sec:astro_obs}) 
are shown in magenta and cyan/green colors, respectively. 
The DD bound from {\sc Damic-M} \cite{DAMIC-M:2025luv} (corresponding to $F(q) = 1$) is 
shown as a black solid line. 
For ID, the darker shaded region corresponds to the variation in the 
galactic DM profile from NFW (benchmark) to a cored-isothermal one. 
On the other hand, for both ID and DD, the light shaded region (bounded by dotted curves) 
corresponds to the variation of $\rho^{\rm DM}_\odot$ in the range [0.2 -- 0.7] $\rm GeV\,cm^{-3}$. 
The black dashed line shows the projection from {\sc Damic-M} (1 kg-yr) \cite{DAMIC-M:2022aks}, 
while the magenta dashed line shows the projection of the MeV telescope {\sc Cosi} 
(1 yr observation), assuming the benchmark NFW profile. 
Note that the {\sc Damic-M} bound and projection are scaled 
from $\rho^{\rm DM}_\odot =$ 0.3 to 0.4 ${\rm GeV \, cm^{-3}}$. 
Other existing exclusions from accelerator searches~\cite{Krnjaic:2022ozp} 
{and DD experiments (based on ER~\cite{DAMIC-M:2025luv} 
and NR~\cite{Chang:2018rso})}, SN1987A~\cite{Chang:2018rso} 
and solar boosted DM search at {\sc PandaX-4T}~\cite{PandaX:2024syk} 
are shown by gray and yellow regions and by the blue line, respectively. 
Note that the accelerator and SN1987A exclusions depend on $\alpha_D$ 
and we adopt $\alpha_D = 0.5$ for both cases. 
The gray dashed-dotted curve shows the bound from the CMB; see the text. 
Finally, the hatched regions show the neutrino floors~\cite{Carew:2023qrj}. 
{\bfseries Right panel}: Similar to the left panel, but under the 
`optimistic approach' specified in sec.~\ref{sec:astro_obs} for the existing 
$X$-ray/$\gamma$-ray observations. 
}
\label{fig:constraints_RV3}
\end{figure*}
We first consider the case in which the dark photon is heavier than the dark matter particle.
In this case, the annihilation of DM proceeds via the so-called 
`Direct Annihilation' process into SM final state particles $f$: $\chi \bar{\chi} \to V^{*}\to f\bar{f}$. 
For definiteness, and to allow a direct comparison with previous literature, we consider the particular case with $m_{V} = 3 \, m_{\rm DM}$. In this case, the annihilation-induced $\gamma$ or $e^\pm$ spectra 
$\left.\frac{dN}{dE}\right\vert_{\rm tot}$ do not depend on the couplings 
$\alpha_D$ or $\epsilon$, and one can derive ID bounds on the total DM pair-annihilation cross-section  
$\langle \sigma v \rangle$ using different astrophysical observations. 

The annihilation cross-section can be expressed as (see, e.g., \cite{LDMX:2018cma,Lin:2022hnt}): 
\begin{equation}
\langle \sigma v \rangle = \langle \sigma v \rangle_{f\bar{f}} \propto \epsilon^2 \, \alpha_D \, \frac{m^2_{\rm DM}}{m^4_{V}} 
= \frac{y}{m^2_{\rm DM}} \, , \hspace{8mm} {\rm with} \hspace{4mm} y = \epsilon^2 \, \alpha_D \, \left(\frac{m_{\rm DM}}{m_{V}}\right)^4 \, .
\label{eq:sigmav_heavyDP}
\end{equation}
Since this scenario corresponds, in the context of DD, to the heavy-mediator case, 
we have $F_{\rm DM}(q)$ = 1. 
By just expressing the quantity $y$ in terms of 
$\bar{\sigma}_e$ (Eq.~(\ref{eq:sigma_e})) as  
\begin{equation}
y = \frac{(q^2_{\rm ref} + m^2_{V})^2 \, m^4_{\rm DM}}{16 \pi \alpha \, m^4_{V} \, \mu^2_{{\rm DM}e}} \, 
\bar{\sigma}_e \, ,
\label{eq:Y_sigmae}
\end{equation}
the connection between the ID and DD observables (respectively $\langle \sigma v \rangle$ and $\bar{\sigma}_e$) is made explicit.

More precisely, the ID bound derived on $\langle \sigma v \rangle$ 
can be translated into the quantity $y$ using: 
$y_{\rm ID} = y_{\rm Th} \, \langle \sigma v \rangle_{\rm ID} / \langle \sigma v \rangle_{\rm Th}$,  
where $y_{\rm Th}$ (corresponding to the thermal freeze-out relic DM) 
can be obtained from \cite{LDMX:2025bog} (as in \cite{Cirelli:2025rky})
and $\langle \sigma v \rangle_{\rm Th} \simeq 2\times10^{-26}$ $\rm cm^3s^{-1}$ is the 
annihilation cross-section for thermal freeze-out relic DM~\cite{Cirelli:2024ssz}. 
This bound on $y$ can then be translated to $\bar{\sigma}_e$ using Eq.~(\ref{eq:Y_sigmae}).

\subsubsection{Results: comparison among different constraints}
\label{sec:results_heavyDP}
In fig. \ref{fig:constraints_RV3}, we compare the DD constraints from {\sc Damic-M} with 
different ID constraints, in the $\bar{\sigma}_e - m_{\rm DM}$ plane related to this DM model. 
The solid lines (for both DD and ID) correspond to the constraints obtained by considering the benchmark NFW 
DM profile with $\rho^{\rm DM}_\odot =$ 0.4 ${\rm GeV\, cm^{-3}}$. 
The existing {\sc Damic-M} upper-limit obtained from \cite{DAMIC-M:2025luv} 
is shown by the black solid line~\footnote{Note that the {\sc Damic-M} bound (and projection) 
shown in this work are rescaled from $\rho^{\rm DM}_\odot =$ 0.3 to 0.4 ${\rm GeV \, cm^{-3}}$, 
as the later value is the benchmark $\rho^{\rm DM}_\odot$ 
considered here, while the bound (projection) provided by the {\sc Damic-M} collaboration 
considered the former value.}. This limit corresponds to the result reported by 
the experimental collaboration based on the search for sub-GeV DM using the {\sc Damic-M} 
prototype detector with an exposure of $\sim 1.3$ kg-day. 

\medskip 
We comment first on the left panel, which presents our main results assuming, for the ID constraints, the `conservative approach' specified in Sec.~\ref{sec:astro_obs}.
The $X$-ray/$\gamma$-ray upper-limit 
is shown by the magenta solid curve. The upper-limits from CR observations 
(obtained as combined bounds from {\sc Voyager-1} and {\sc Ams-02} data) are shown by cyan and green 
solid curves, corresponding to the galactic propagation models `prop.~a' and `prop.~c' 
mentioned in Sec.~\ref{sec:DM_CR_signal}, respectively. 
The model `prop.~c', which features a comparatively 
larger reacceleration, provides a significantly stronger bound for a $m_{\rm DM} \lesssim 100$ MeV.  
The inclusion of {\sc Ams-02} data helps in improving the bounds for a comparatively heavier DM 
(above a few hundreds of MeV).  

As it can be seen from this figure, in almost all cases, apart from the CR constraints 
for $m_{\rm DM} \lesssim 10$ MeV corresponding to a propagation model with a small reacceleration effect, 
the ID bounds from existing $X$-ray/$\gamma$-ray and CR observations dominate over the present 
{\sc Damic-M} limit, considering a standard NFW profile for the galactic DM distribution.  

In addition to the benchmark NFW DM profile (which is cuspy), we also consider a 
cored isothermal profile, Eq.~(\ref{eq:rho_Iso}) 
(with the same value of $\rho^{\rm DM}_\odot = 0.4$ $\rm GeV\,cm^{-3}$), 
for which the corresponding weakening of existing ID constraints are highlighted in 
fig.~\ref{fig:constraints_RV3} by the dark colored bands. Considering this, 
the {\sc Damic-M} upper-bound (which is unaltered under this change) can provide a 
slightly better constraint compared to, e.g., the $X$-ray/$\gamma$-ray limit, for 
a DM mass around 10 MeV. 

As mentioned in Sec~\ref{sec:formalism_ID}, we also consider a variation of the galactic (NFW) DM profile 
based on the variation of $\rho^{\rm DM}_\odot$ in the range [0.2 -- 0.7] $\rm GeV\,cm^{-3}$. 
The corresponding variations in the {\sc Damic-M} and ID upper-limits 
are indicated in fig.~\ref{fig:constraints_RV3} by the light colored bands (bounded by thin dotted lines)
around the solid curves. 
For DD experiments like {\sc Damic-M}, where the DM signal is proportional to $\rho^{\rm DM}_\odot$, 
the thickness of this band is just equal to the factor $0.7/0.2 = 3.5$. 
On the other hand, for ID searches this variation in the DM density shows a significant impact, 
mainly due to: (i) the scaling of the annihilation signal with the square of the DM density, 
and (ii) the overall spatial shape of the DM profile which is remodeled in order to keep the quantity $M_{\rm DM} (r\le200\,{\rm kpc})$ a constant (see Sec~\ref{sec:formalism_ID}). 
From fig.~\ref{fig:constraints_RV3} we can see that 
considering such a variation, {\sc Damic-M} can provide a somewhat better constraint in the 
range $3\,{\rm MeV} \lesssim m_{\rm DM} \lesssim 30\,{\rm MeV}$. 

\medskip

In the right panel of fig.~\ref{fig:constraints_RV3} we show 
the ID upper-limits obtained using the `optimistic approach', which takes into account 
the standard astrophysical backgrounds (as mentioned in Sec.~\ref{sec:astro_obs}). 
Here only the $X$-ray/$\gamma$-ray observations are considered and the color codes are the same 
as the ones shown in the left panel of the figure. As expected, in this case the ID bounds from existing data can be improved noticeably, 
leaving a very narrow mass window ($m_{\rm DM} \simeq 5$ MeV) where {\sc Damic-M} 
can dominate over ID searches even after considering the full variation due to 
the DM profile uncertainties. We note in passing that the {\sc Damic-M} constraints obtained 
in \cite{DAMIC-M:2025luv} are derived taking into account a possible DD backgrounds, hence they are in principle consistent with the `optimistic' approach. 

\bigskip

Next, we show in fig.~\ref{fig:constraints_RV3} the future prospect of 
{\sc Damic-M} (shown by the black dashed line) in probing the sub-GeV DM-electron interaction 
and that of the upcoming space-based MeV telescope {\sc Cosi} (the magenta dashed line). 
The {\sc Damic-M} sensitivity projection is obtained from \cite{DAMIC-M:2022aks} 
and corresponds to a target exposure of 1 kg-year. On the other hand the 
prospect of the {\sc Cosi} telescope is obtained as mentioned in Sec.~\ref{sec:astro_obs} 
for an observation time of 1 year. Here we only consider the benchmark NFW DM profile 
with the benchmark value of $\rho^{\rm DM}_\odot$. 
This comparison shows that, with the targeted exposure, {\sc Damic-M} will be 
able to provide a significantly better sensitivity than {\sc Cosi}, for a DM mass in the range 2 -- 80 MeV. 
On the other hand, below and above this mass range {\sc Cosi} will provide 
a much better probe. Thus these two instruments should complement each other 
in exploring the sub-GeV mass range of the DM-electron interaction under the considered model. 

\bigskip

Finally, for comparison, we also show in fig.~\ref{fig:constraints_RV3} other existing exclusions in the same 
parameter space coming from: 
(1) solar boosted DM search at {\sc PandaX-4T} (the blue line)~\cite{PandaX:2024syk}, 
(2) accelerator searches~\cite{Krnjaic:2022ozp} {and other DD experiments 
based on electron-recoil (ER)~\cite{DAMIC-M:2025luv} and 
nuclear recoil (NR)~\cite{Chang:2018rso}} (the gray region), 
and (3) SN1987A observation (the yellow region)~\footnote{\label{footnote:SNbounds}{Note that 
the SN exclusion is shown here for the sake of comparison with the DD and ID bounds
and is adopted directly from \cite{Chang:2018rso}.
The treatments/approximations used to obtain such an
exclusion are improved in more recent studies \cite{Caputo:2022rca, Fiorillo:2025yzf},
overall leading to a weaker exclusion from SN1987A.
This is even more significant when $\alpha_D$ is large enough
(as adopted in this figure). In this case,
due to the inclusion of self-interactions, the ceiling of the
SN exclusion (in the $\bar{\sigma}_e - m_{\rm DM}$ plane)
is expected to move down by almost two orders of magnitude, 
allowing more parameter space. See \cite{Fiorillo:2024upk}
for improved discussions along this line (in the context of
millicharged particles).}}~\cite{Chang:2018rso}. 
Note that the accelerator and SN1987A exclusions depend on $\alpha_D$ 
(we adopt $\alpha_D = 0.5$ for both cases), while the ID and 
DD bounds (like {\sc Damic-M}) are independent of $\alpha_D$, as discussed above. 
We also highlight with hatched regions the neutrino floors 
(for silicon and liquid xenon DD experiment) from \cite{Carew:2023qrj} 
(see also \cite{Cirelli:2024ssz}) for the considered model. 
As it can be seen, compared to these other searches, 
the current {\sc Damic-M} bound excludes some more parameter space 
for a DM mass smaller than $\sim$30 MeV. On the other hand, the existing ID searches 
constrain a larger parameter space for almost the entire DM mass range. 
Note that, for $m_{\rm DM}$ larger than a few hundreds of MeV, the ID bounds can already 
reach the level of the neutrino floor or deeper, depending on the astrophysical uncertainties. 
Both the upcoming MeV telescopes and the future {\sc Damic-M} will 
improve their bounds even further 
as discussed above. For example, {\sc Cosi} can in principle probe the DM-$e$ interaction 
deep inside the $\nu$-floor for $m_{\rm DM} \gtrsim 150$ MeV. 

\bigskip

In addition to the bounds discussed above, we also show in fig.~\ref{fig:constraints_RV3} for comparison the constraint on annihilating DM 
that can come from the CMB observation (the gray dashed-dotted curve). 
This bound for the considered model 
is estimated using the efficiency factors for $e^\pm$ and $\gamma$ from \cite{Slatyer:2015jla}, 
and following the same methodology used in \cite{Cirelli:2025rky}. 
As it can be seen from the figure, under the `conservative approach', considering 
the uncertainty related to the DM profile, existing $X$-ray/$\gamma$-ray 
observations can constrain a new parameter space allowed by the CMB 
for $m_{\rm DM} \gtrsim 100$ MeV. On the other hand, under the 
`optimistic approach', the $X$-ray/$\gamma$-ray exclusion can reach at the level 
of CMB bound for $m_{\rm DM} \gtrsim 150$ MeV even considering the standard NFW profile. 
In the future, {\sc Damic-M} will be capable of constraining new parameter space 
for DM-$e$ interaction below the CMB limit for $2\,{\rm MeV} \lesssim m_{\rm DM} \lesssim 40\,{\rm MeV}$, while {\sc Cosi} will be able to probe a new portion of the parameter space, for $m_{\rm DM} \gtrsim 100$ MeV.

Note that, unlike the usual indirect detection searches which 
are based on the observations of DM signal produced in the galactic systems in the present Universe, the above-mentioned CMB bound is based on the cosmological effects of DM annihilation 
in the early universe near the epoch of CMB. 
Thus such a bound can be evaded in a number of ways considering 
different scenarios. One possibility can be a scenario where the 
DM candidate is partially produced through the decay of a heavier 
dark sector particle after recombination (see, e.g., \cite{DEramo:2018khz}). 
Another possibility can be a scenario of inelastic DM where the DM candidate is accompanied by 
another slightly heavier dark sector particle. In this case the DM annihilation in the early 
Universe can be suppressed significantly (required to evade the CMB bound), 
while it becomes important in the present day galaxies to produce detectable signals 
in the $\gamma$-ray telescopes (see, e.g., \cite{Berlin:2023qco}). 
Another (fine-tuned) possibility could be a resonant DM annihilation 
(via a mediator $m_V \simeq 2 \, m_{\rm DM}$) which is enhanced in the present day 
galactic environment due to the high dispersion velocity of DM, 
but suppressed during the CMB epoch.

\subsection{`Light' dark photon: $m_{V} < m_{\rm DM}$}
\label{sec:light_DP}
We now move to the case in which the dark photon is lighter than the DM particle: $m_{V} < m_{\rm DM}$.  
In the case, the so-called `Secluded Annihilation' channel $\chi \bar{\chi} \to V V\to f\bar{f} f'\bar{f'}$ is kinematically allowed and opens up.  The total DM pair-annihilation cross-section $\langle \sigma v \rangle$ is thus given by
\begin{equation}
\langle \sigma v \rangle = \langle \sigma v \rangle_{f\bar{f}} + \langle \sigma v \rangle_{VV},
\end{equation} 
with (see, e.g., \cite{Lin:2022hnt})
\begin{equation}
\langle \sigma v \rangle_{f\bar{f}} \propto \epsilon^2 \, \alpha_D \, \frac{\pi \alpha_f}{m^2_{\rm DM}}
\end{equation}
and (see, e.g., \cite{Liu:2014cma, Green:2018qwo})
\begin{equation}
\langle \sigma v \rangle_{VV} \simeq \frac{\pi \alpha^2_D}{m^2_{\rm DM}} \, 
\frac{\left( 1 - m^2_{V}/m^2_{\rm DM} \right)^{3/2}}{\left( 1 - m^2_{V}/2 m^2_{\rm DM} \right)^2}. 
\label{eq:sv_VV}
\end{equation}
The latter usually dominates the DM annihilation process, leading to 
\begin{equation}
\langle \sigma v \rangle \simeq \langle \sigma v \rangle_{VV} \, .
\label{eq:sv_light_med}
\end{equation}
We here fix the case $m_{V} = m_{\rm DM} \, / \, 3$ and thus, 
for $1\,{\rm MeV} \le m_{\rm DM} \le 1\,{\rm GeV}$, we are still in the heavy mediator case for DD.

\medskip

\begin{figure*}[!t]
\begin{minipage}{0.60\textwidth}
\hspace{-0.5cm}
\includegraphics[width=\textwidth]{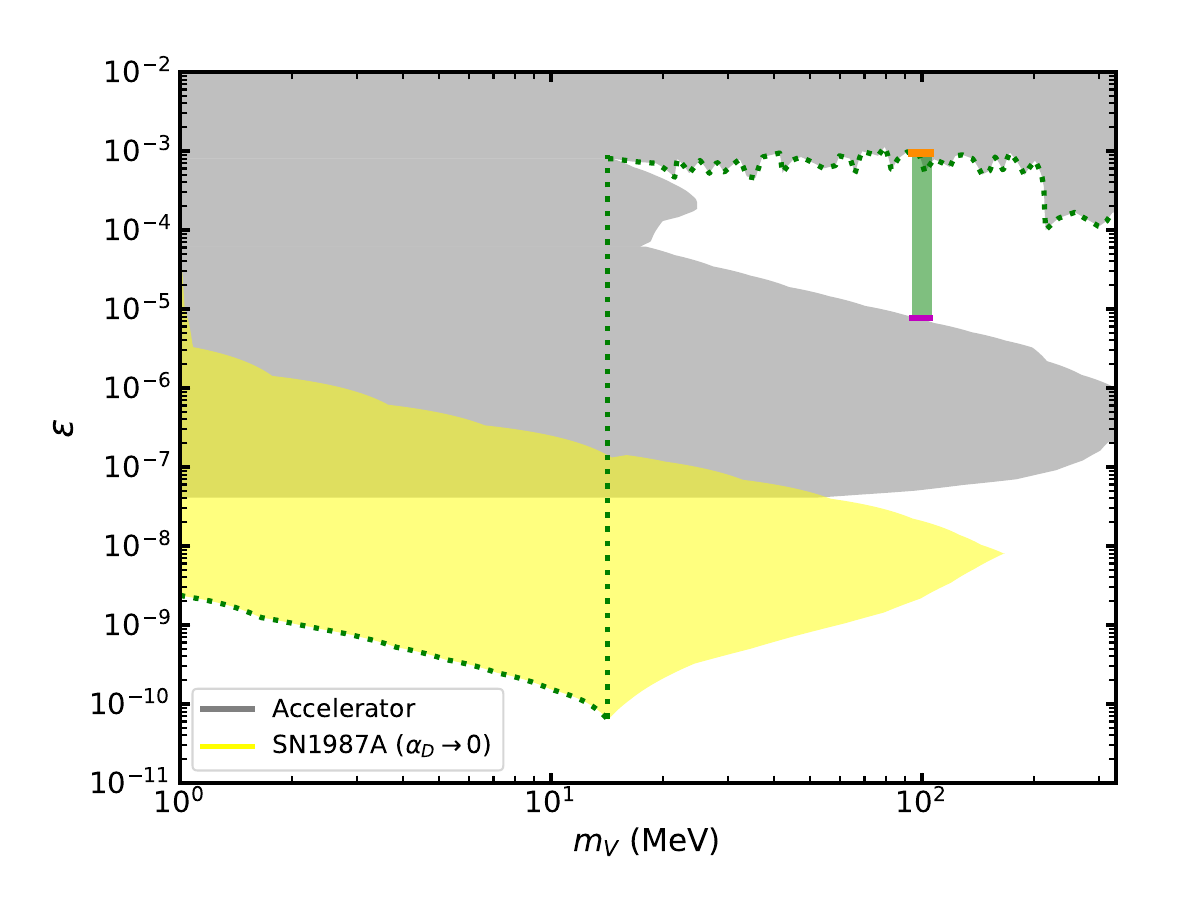}\\
\end{minipage}
\begin{minipage}{0.38\textwidth}
\vspace{-12mm}
\caption{\em {\bfseries Exclusions in the $\epsilon - m_V$ plane} from accelerator searches 
\cite{Batell:2022dpx} and SN1987A \cite{Chang:2018rso}. 
The SN excluded area is reported for $\alpha_D \to 0$ (see the text); the accelerator one does not depend on $\alpha_D$. 
The green dotted line indicates the maximum limit of $\epsilon$ 
(as a function of $m_V$) allowed when combining the observations. 
The green strip identifies the values of $\alpha_{\rm D}$ that 
map onto the green dot in the left panel of fig.~\ref{fig:constraints_RV0.33} (see the text).
}
\label{fig:constraints_epsilon}
\end{minipage}
\end{figure*}

Using the astrophysical observations one can obtain the ID bounds on 
$\langle \sigma v \rangle$. Then such bounds can first be translated to $\alpha_D$ 
using Eqs.~(\ref{eq:sv_VV}) and (\ref{eq:sv_light_med}), and then to $\bar{\sigma}_e$ using 
Eq.~(\ref{eq:sigma_e}), with an additional information on $\epsilon$. Since $\bar{\sigma}_e$ 
is proportional to $\epsilon^2$ (Eq.~\ref{eq:sigma_e}), 
we choose, for the above translation, the maximum limit set on $\epsilon$ 
by independent experimental observations.  
This is illustrated in fig.~\ref{fig:constraints_epsilon}, where we show 
the exclusions in the $\epsilon - m_V$ plane from accelerator searches~\cite{Batell:2022dpx} 
and SN1987A~\cite{Chang:2018rso}. 
Note that the exclusion region from SN1987A depends on $\alpha_D$ and actually moves slightly towards lower $\epsilon$ with increasing $\alpha_D$. We take the limit $\alpha_D \rightarrow 0$ case, since $\alpha_D$ is constrained to be small from ID bounds 
(see figure \ref{fig:sv_constraints_RV0.33} below). 
On the other hand, the accelerator exclusion is independent of $\alpha_D$.
The green dotted line indicates the maximum limit of $\epsilon$ 
(as a function of $m_V$) allowed by the observations. 

\medskip

We checked that, for this scenario ($m_{V} = m_{\rm DM} \, / \, 3$), too, 
the DM induced spectrum $\left.\frac{dN}{dE}\right\vert_{\rm tot}$ and hence 
the ID bound obtained on $\langle \sigma v \rangle$ do not depend much on 
$\alpha_D$ (unless it is smaller than the range studied here) 
nor on $\epsilon$ (unless it is very large compared to the experimental upper-limit). 
Also, for the range of $\epsilon$ considered here, the decay-length of MeV-GeV 
dark photons is much smaller than a parsec, leading to an effectively instantaneous 
injection of $e^\pm$ pairs from their decays (which is required to produce the ID signals within the observational region of the Galaxy).

\subsubsection{Results: comparison among different constraints}

\begin{figure*}[!t]
\begin{tabular}{cc}
\hspace{-9mm} \includegraphics[width=0.53\textwidth]{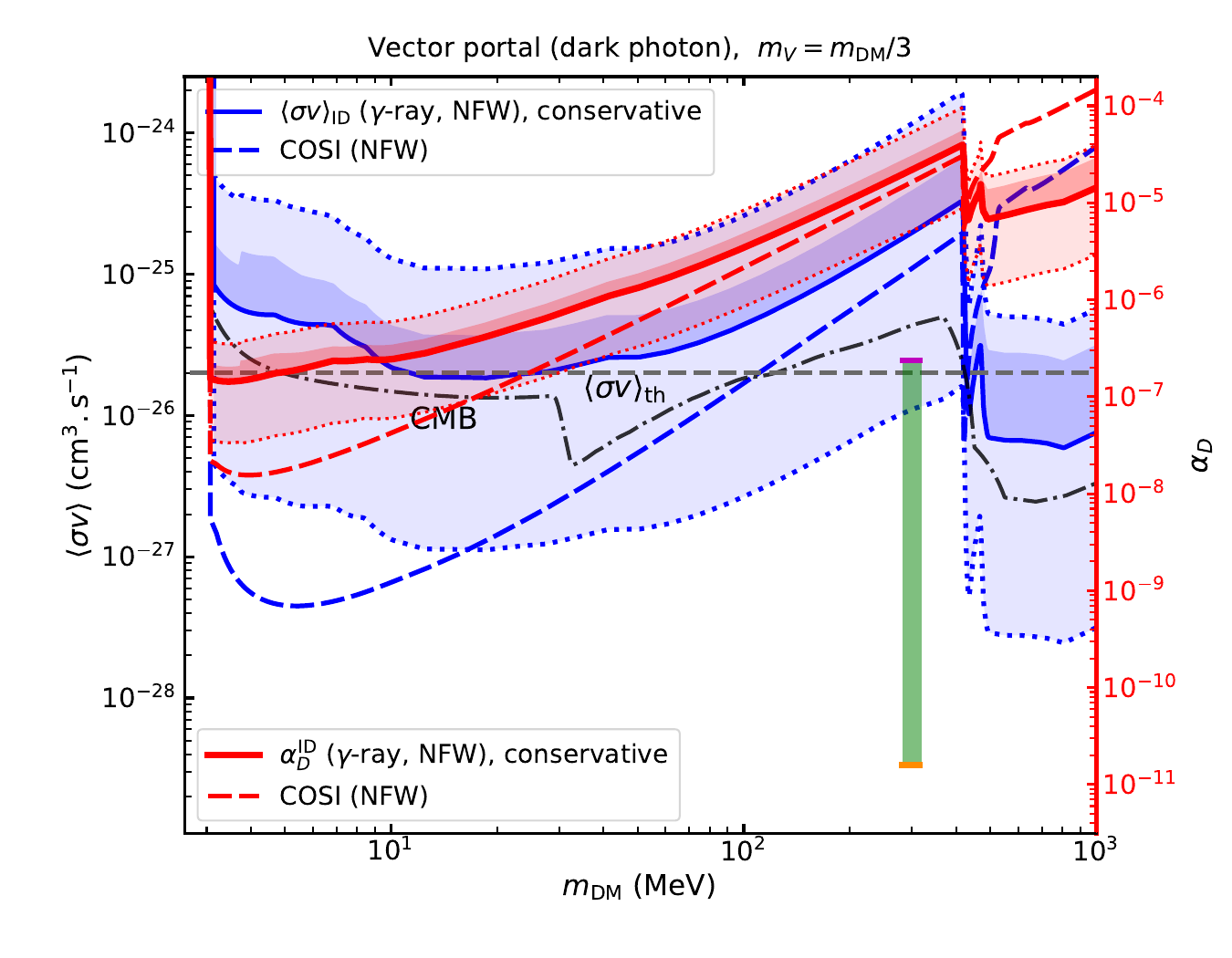}\hspace{-1mm}
\includegraphics[width=0.53\textwidth]{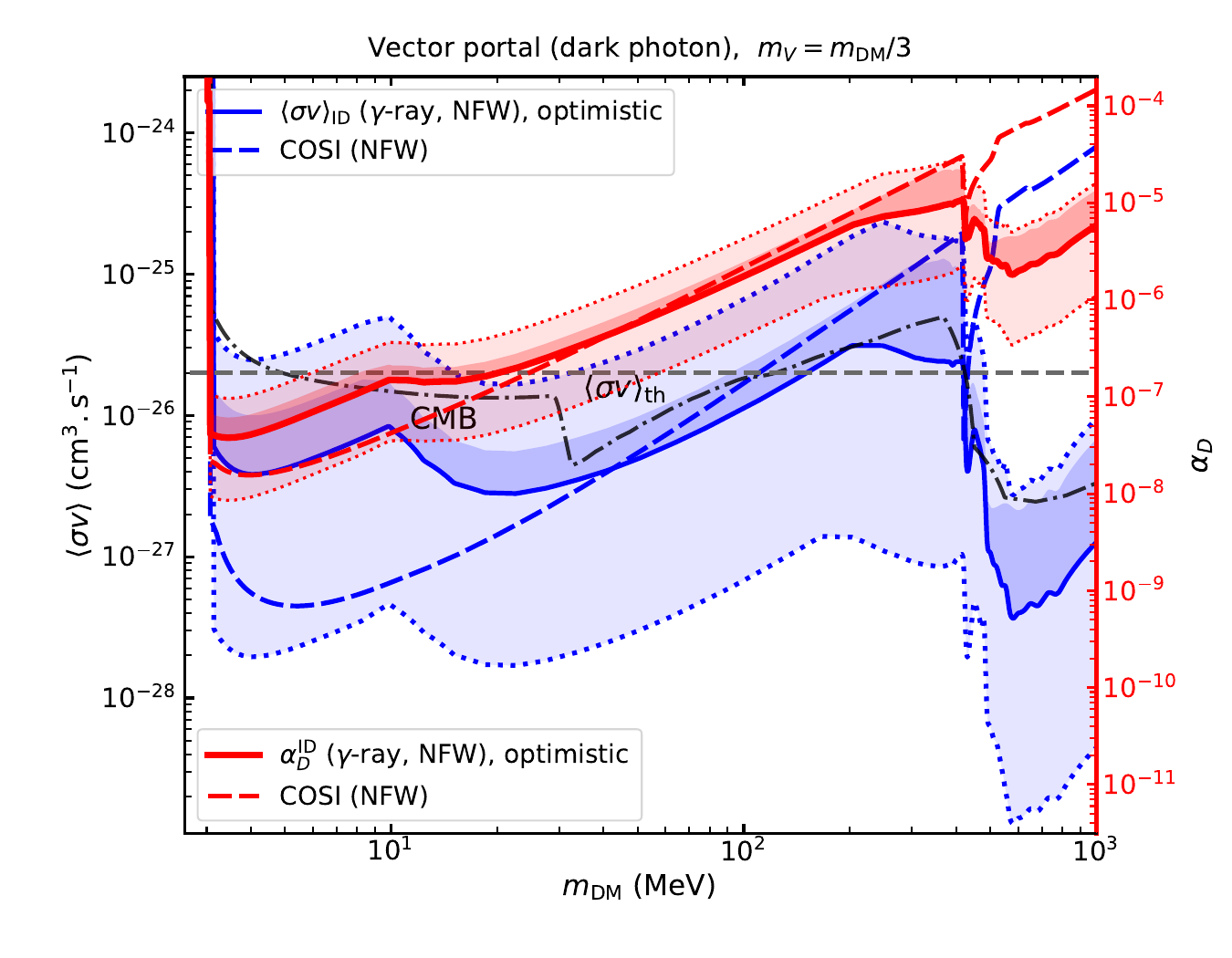}\\
\end{tabular}
\caption{\em {\bfseries ID bounds} (from existing $X$-ray/$\gamma$-ray observations) on 
$\langle \sigma v \rangle$ and equivalently on $\alpha_D$. 
The dark shaded regions correspond to the variation in the 
DM profile from NFW to the cored-Isothermal one, while the light shaded regions 
(bounded by dotted curves) correspond to the variation of $\rho^{\rm DM}_\odot$ 
in the range [0.2 -- 0.7] $\rm GeV\,cm^{-3}$. 
The dashed lines show the projection of {\sc Cosi} (1 yr observation). 
The gray dashed-dotted curve show the bound from CMB, see the text. The gray 
dashed line indicates the thermal $\langle \sigma v \rangle$.
{\bfseries Left}: Considering the `conservative approach' defined in sec.~\ref{sec:astro_obs}. The green strip identifies the values of $\alpha_{\rm D}$ that map onto the green dot in the left panel of fig.~\ref{fig:constraints_RV0.33} (see the text).
{\bfseries Right}: Considering the `optimistic approach'.
} 
\label{fig:sv_constraints_RV0.33}
\end{figure*}

In fig.~\ref{fig:sv_constraints_RV0.33} 
we present the limits for this DM model from existing $X$-rays/$\gamma$-rays observations 
considering the `conservative approach' (left panel, our main results) 
as well as the `optimistic approach' (right panel). 
Here we show only $X$-ray/$\gamma$-ray constraints, since the bounds from 
CR observations turn out to be comparatively much weaker. 
The solid blue curves show the limits on $\langle \sigma v \rangle$ considering 
the benchmark NFW profile, while the solid red lines 
are the corresponding limits on the coupling $\alpha_D$. 
Like in fig.~\ref{fig:constraints_RV3}, the darker shaded regions correspond to 
the variation in the DM profile from NFW to the cored-isothermal one, 
while the light shaded regions (bounded by dotted curves) 
correspond to the variation of $\rho^{\rm DM}_\odot$ 
in the interval [0.2 -- 0.7] $\rm GeV\,cm^{-3}$. The blue (red) dashed curve shows 
the projection of the upcoming MeV telescope {\sc Cosi} (for 1 yr of observation) 
in probing $\langle \sigma v \rangle$ ($\alpha_D$) considering the benchmark NFW profile. 
For comparison, we show bound from CMB (the gray dash-dotted curve) which for this 
scenario is computed in the same way mentioned in Sec.~\ref{sec:results_heavyDP}. 
In addition, we also indicate the thermal relic annihilation cross section by the gray dashed line. 
Note that the cut off of all ID bounds at $m_{\rm DM} \simeq 3$ MeV (left edge of the plots) is due to the 
fact that the case $m_{\rm DM} < 3$ MeV corresponds to a dark photon lighter than an MeV which 
cannot decay into any charge particle pair to produce an ID signal. 
The abrupt kink at $m_{\rm DM} \gtrsim 400$ MeV corresponds, on the other hand, to the opening of dark photon decays into pions.

\medskip

As it can be seen from fig.~\ref{fig:sv_constraints_RV0.33}, 
the bounds from existing $X$-ray/$\gamma$-ray already 
reach at the level of the CMB bound under the conservative approach and 
can overcome the CMB bound considering the variation in the DM profile. 
Taking the optimistic approach with the standard astrophysical $X$-ray/$\gamma$-ray background, 
it is possible to overcome the CMB bound at almost all DM masses in the range MeV -- GeV. 
Ultimately, {\sc Cosi} will be able to provide a significant improvement in probing 
the parameter space for $m_{\rm DM} \lesssim 400$ MeV, 
and should easily overcome the CMB bound for $m_{\rm DM} \lesssim 100$ MeV. 

\bigskip

\begin{figure*}[!t]
\begin{tabular}{cc}
\hspace{-18mm} \includegraphics[width=0.6\textwidth]{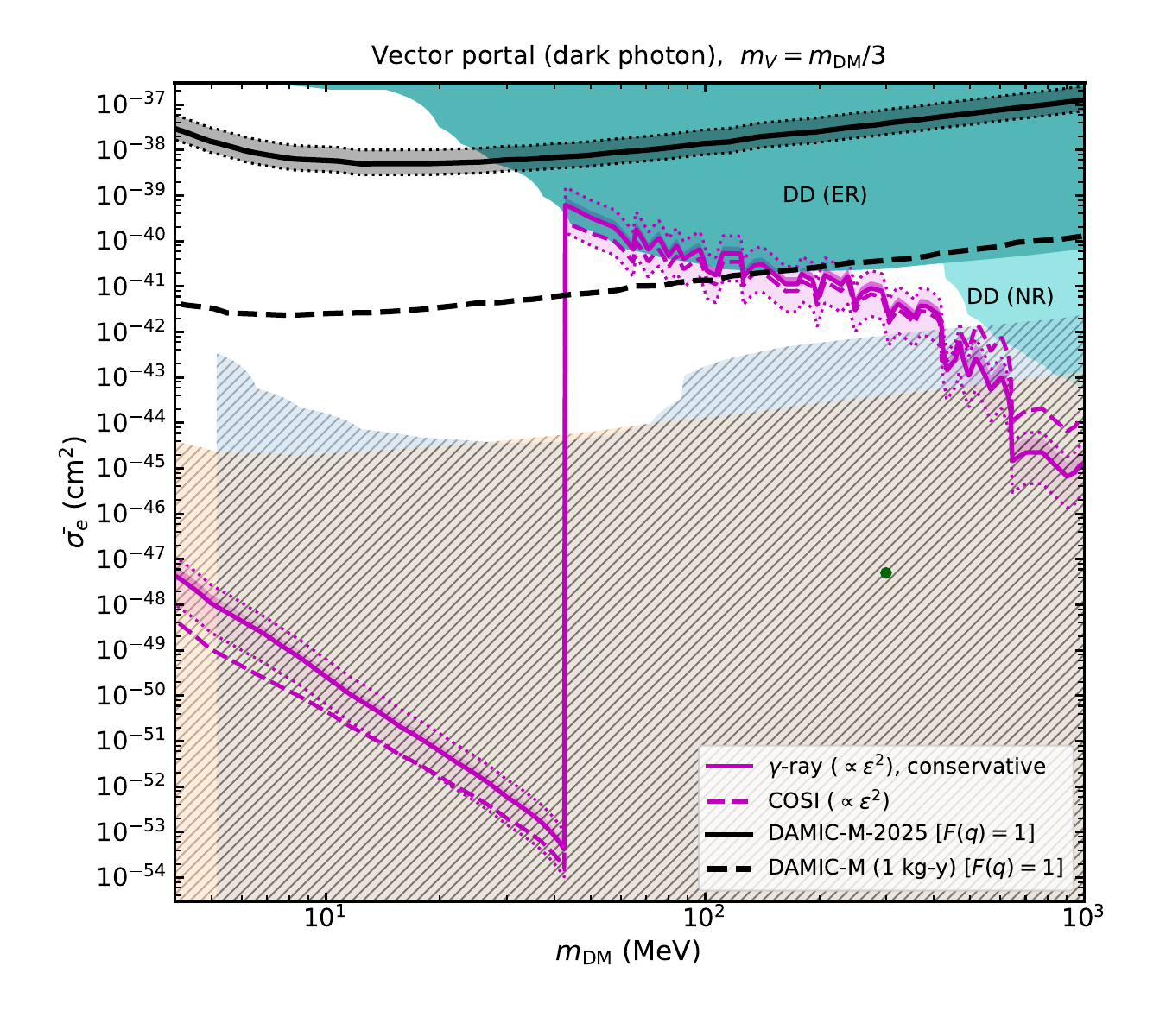} & \hspace{-10mm}
\includegraphics[width=0.6\textwidth]{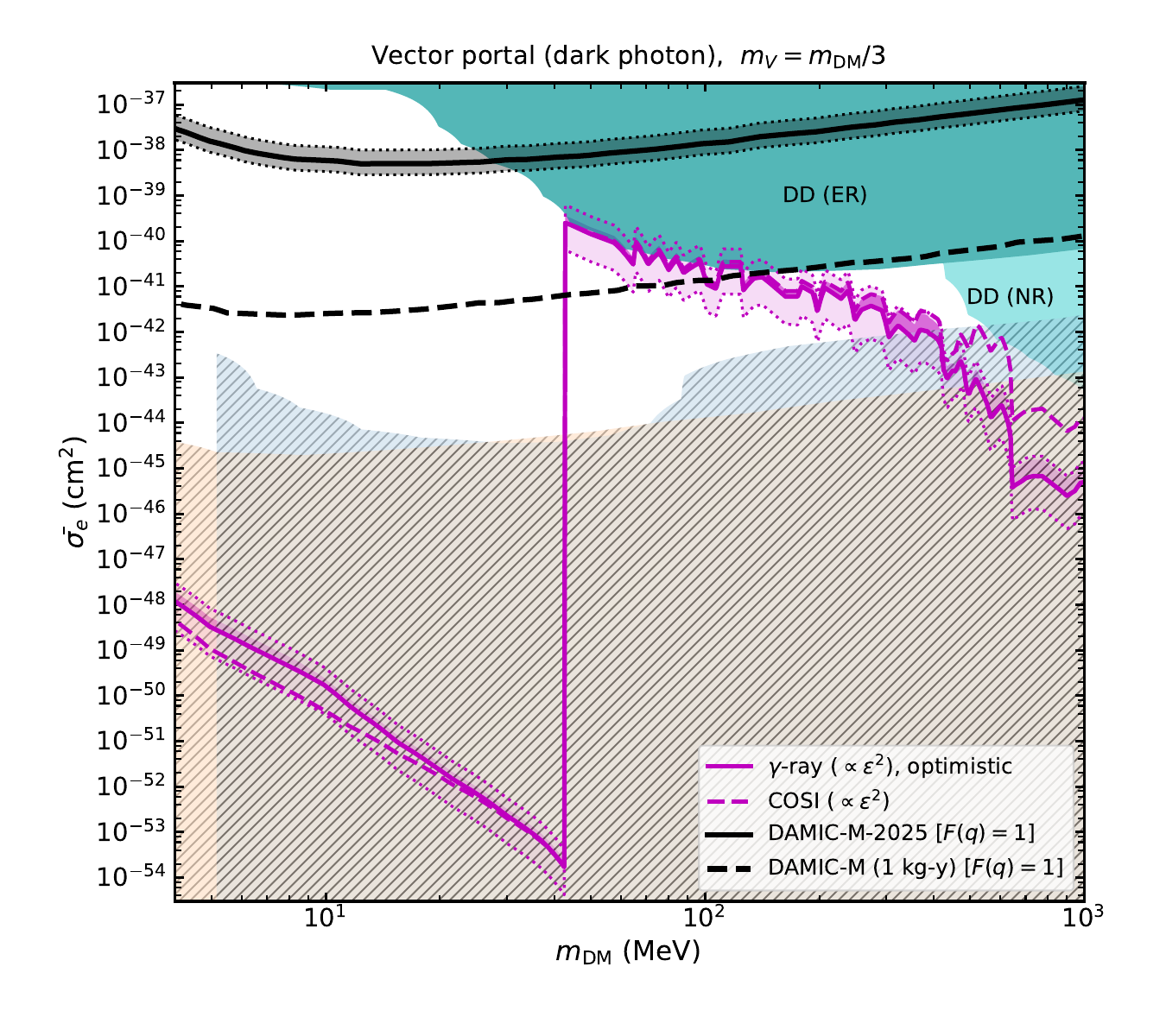}\\
\end{tabular}
\vspace{-3mm}
\caption{\em Comparisons of ID and DD constraints for the {\bfseries vector portal model}, with a {\bfseries light dark photon}, in the $\bar{\sigma}_e - m_{\rm DM}$ plane.
The magenta lines correspond to the upper-limits on $\bar{\sigma}_e$ obtained as the 
combination of the ID upper bound on $\alpha_D$ (the red lines in fig.~\ref{fig:sv_constraints_RV0.33}) 
and the maximum limit on $\epsilon$ 
(the green dotted line of fig.~\ref{fig:constraints_epsilon}). 
The {\sc Damic-M} bound from \cite{DAMIC-M:2025luv} (for $F(q) = 1$) 
is shown by the black solid line. 
The meaning of the shaded bands for both {\sc Damic-M} and ID contraints is the same as in fig.~\ref{fig:constraints_RV3} and fig.~\ref{fig:sv_constraints_RV0.33}. 
The black dashed line indicates the projections for {\sc Damic-M} 
(with an exposure of 1 kg-yr) \cite{DAMIC-M:2022aks}, while 
the magenta dashed line shows the {\sc Cosi} projection (for 1 yr observation). 
The existing combined exclusions from other DD experiments are shown 
by dark cyan (ER experiments) \cite{DAMIC-M:2025luv} 
and light cyan (NR experiments) \cite{Chang:2018rso} shaded areas. 
The hatched regions show the neutrino floors. 
{\bfseries Left panel}: Considering the `conservative approach' for ID, as defined in sec.~\ref{sec:astro_obs}. The green dot identifies an example of a point allowed by current constraints, which would be however excluded by a different choice of the $\alpha_{\rm D}$ and $\epsilon$ parameters (see the text). 
{\bfseries Right panel}: Considering the `optimistic approach' for ID.} 
\label{fig:constraints_RV0.33}
\end{figure*}

Next, in fig.~\ref{fig:constraints_RV0.33} we place the ID upper-bounds and projection 
(shown by the magenta curves) in the $\bar{\sigma}_e - m_{\rm DM}$ plane and compare them with 
the DD upper-bounds and projection from {\sc Damic-M} (black curves). 
The ID constraints on $\bar{\sigma}_e$ are obtained as the 
combination of the ID bounds on $\alpha_D$ and the maximum experimental limit on $\epsilon$. 
The line styles and the representations of the shaded bands for both {\sc Damic-M} and ID 
are the same as the ones shown in fig.~\ref{fig:constraints_RV3}. 
For comparison, we also show in this plane the exclusions from other DD 
experiments based on electron-recoil (ER) (taken from \cite{DAMIC-M:2025luv}) 
and nuclear recoil (NR) (taken from \cite{Chang:2018rso}). The hatched region 
shows the neutrino floors for this model. In this plane we do not include the 
CMB bound (obtained in fig.~\ref{fig:sv_constraints_RV0.33}) 
as they are at the level of (or under some circumstances 
even weaker than) the existing ID bounds based on $X$-rays/$\gamma$-rays.   

\medskip

As one can see from fig.~\ref{fig:constraints_RV0.33}, 
for the present model, the current ID upper-bound is much more stringent than 
the present {\sc Damic-M} limit for the entire DM mass range of interest. 
In addition, these current ID 
observations also provide a significantly better probe in the DM-e parameter space 
compared to other DD experiments (even those based on the nuclear recoil) 
for sub-GeV DM with masses above $\sim200$ MeV and below $\sim40$ MeV. 
In fact, the ID observations exclude a large amount of parameter space 
in the $\bar{\sigma}_e - m_{\rm DM}$ plane which is not accessible to DDs 
because of the neutrino floor. Note that, the sharp drop in the ID limit in this plane 
for $m_{\rm DM} \lesssim 40$ MeV is due to the sharp drop of the maximum 
upper-limit on $\epsilon$ (see the green dotted line in the bottom panel of fig.~\ref{fig:sv_constraints_RV0.33}). 

Regarding the ID and DD projections we see that {\sc Cosi} will improve the current ID bounds for $m_{\rm DM} \lesssim 100$ MeV (as it is also evident in fig.~\ref{fig:sv_constraints_RV0.33}), while the future {\sc Damic-M} \cite{DAMIC-M:2022aks}  is expected to be more constraining than {\sc Cosi} only in the $\sim 40 - 130$ MeV mass range.

\bigskip

We end this section with a side remark on the procedure employed to derive the constraints. 
We note that, in the $\bar{\sigma_e} - m_{\rm DM}$ plane, the ID bounds are proportional to $\epsilon^2$, while 
the DD bounds are independent of $\epsilon$. 
Hence, the regions below the red lines in fig.~\ref{fig:constraints_RV0.33} are allowed/disallowed depending on the choices of $\alpha_D$ and $\epsilon$.
We show as allowed any point in the plane for which at least one pair $\{\alpha_{\rm D},\epsilon\}$ of allowed points exists. 
For instance, the point identified by the green dot in the left panel of fig.~\ref{fig:constraints_RV0.33} is allowed as it maps onto a pair of points within the green strips of fig.~\ref{fig:sv_constraints_RV0.33} and \ref{fig:constraints_epsilon}, which lie in the respective allowed areas. 
The same green dot in $\bar\sigma_e$ can of course be reproduced by other choices of pairs of values. For instance, if we choose a value of $\alpha_{\rm D}$ which lies below the ID bound of fig.~\ref{fig:sv_constraints_RV0.33} (hence, allowed by ID) but above the upper edge of the green strip, the corresponding value of $\epsilon$ needed to reproduce the green dot would lie below the lower edge of the green strip in fig.~\ref{fig:constraints_epsilon}, hence excluded by accelerator and/or supernova constraints. This would put the green dot in an excluded area,  incorrectly.
The procedure that we follow, instead, takes into account the interplay between the bounds on $\epsilon$ and on $\alpha_{\rm D}$, and correctly identifies all the excluded areas of the $\bar{\sigma_e} - m_{\rm DM}$ plane. 
Note that, in the `heavy dark photon' case of sec.~\ref{sec:heavy_DP}, this caution does not apply: both the ID and DD observables carry a dependence on $\epsilon^2 \times \alpha_{\rm D}$ (see eqs.~(\ref{eq:sigma_e}) and (\ref{eq:sigmav_heavyDP})) hence a bound on $\langle \sigma v \rangle$ is directly translated into a bound on $\bar\sigma_e$ and vice-versa, as encoded by the quantity $y$ of eq.~(\ref{eq:Y_sigmae}).

\subsection{`Very light' dark photon: $m_{V} \ll 1$ keV}
\label{sec:Very_light_DP}

\begin{figure*}[!t]
\centering
\includegraphics[width=0.6\textwidth]{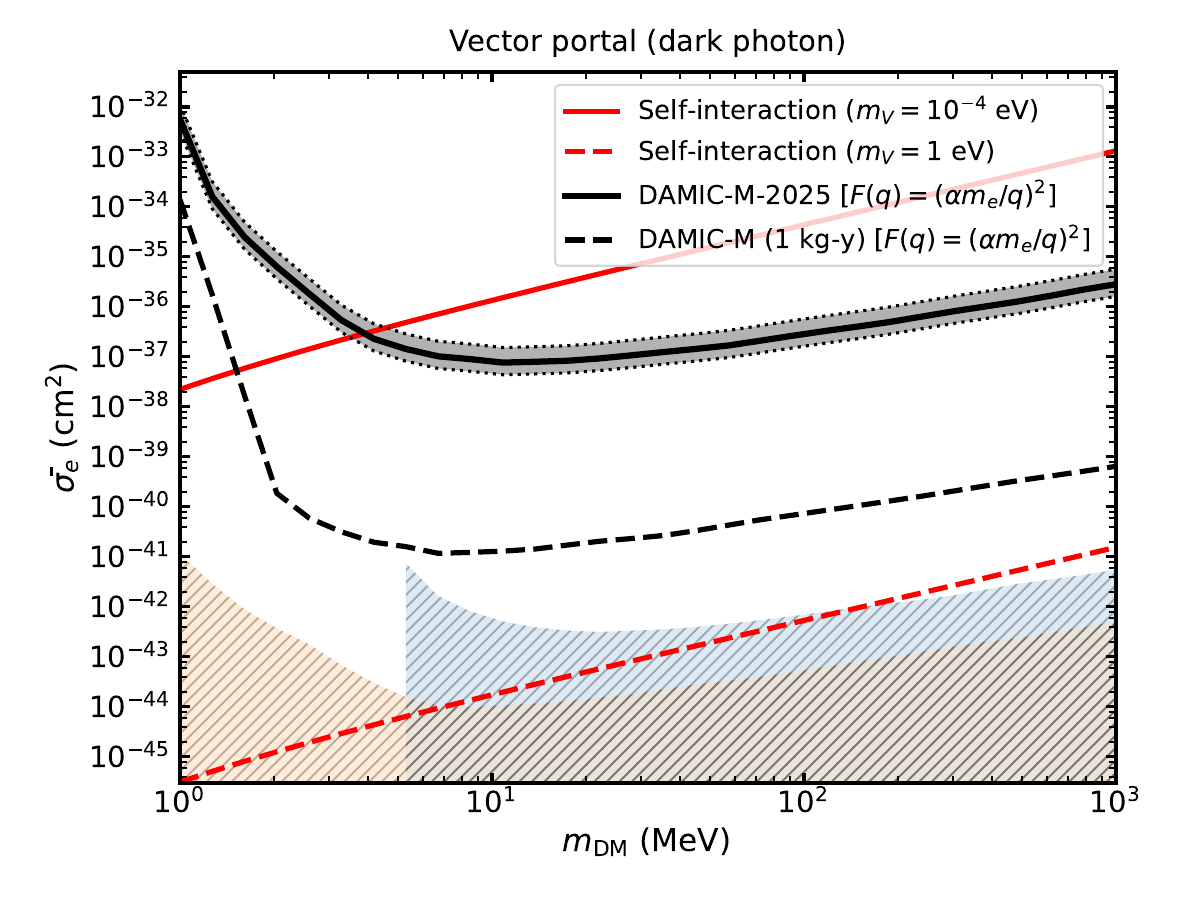}\\
\vspace{-3mm}
\caption{\em Constraints for the {\bfseries vector portal model}, in the case of a {\bfseries very light dark photon}, in the $\bar{\sigma}_e - m_{\rm DM}$ plane.
The {\sc Damic-M} bound from \cite{DAMIC-M:2025luv} 
(with $F(q) = (\alpha m_e / q)^2$) is 
shown by the black line. The gray band corresponds to the 
variation of $\rho^{\rm DM}_\odot$ in the range [0.2 -- 0.7] $\rm GeV\,cm^{-3}$. 
The black dashed line shows the {\sc Damic-M} (1 kg-yr) 
projection from \cite{DAMIC-M:2022aks}. 
Note that the {\sc Damic-M} bound and projection are scaled from 
$\rho^{\rm DM}_\odot =$ 0.3 to 0.4 ${\rm GeV \, cm^{-3}}$. 
The red lines correspond to the DM self-interaction bounds for two values of $m_V$:  
$10^{-4}$ eV (solid) and 1 eV (dashed). The hatched regions 
show the neutrino floors.}
\label{fig:constraints_mVultralight}
\end{figure*}

We also consider, for completeness, another scenario where $m_{V} \ll 1$ keV. 
In this case, the computation of the total DM pair-annihilation cross-section 
$\langle \sigma v \rangle$ in terms of the model parameters follows the same 
expressions described in sec.~\ref{sec:light_DP}. 
However, since the process $\chi \bar{\chi} \longrightarrow VV$ dominates 
the DM annihilation and $m_{V} \ll 1$ MeV (i.e., the dark photon mass is smaller than 
any charged SM fermion), the production of 
SM particles in the final state of DM annihilation is very much suppressed. 
Thus in this case, the ID bounds derived using the standard processes discussed in 
sec.~\ref{sec:formalism_ID} are effectively negligible. For this type 
of scenario, DD experiments like {\sc Damic-M} provide instead the leading constraints 
in probing the DM interactions. 

\medskip

In fig.~\ref{fig:constraints_mVultralight} we show the {\sc Damic-M} upper-limit and the projection 
with black lines. 
Note that, for this case of very light mediator, the pertinent form factor used for bounds and projections 
is $F(q) = (\alpha m_e / q)^2$. The band around the {\sc Damic-M} bound corresponds to the variation of $\rho^{\rm DM}_\odot$ 
in the range [0.2 -- 0.7] $\rm GeV\,cm^{-3}$. 

On the other hand, in this very light dark photon case a strong bound 
(on $\alpha_D$) arises from DM self-interaction (SIDM)~\cite{Knapen:2017xzo}. This is expressed as: 
\begin{equation}
\frac{\sigma_{\rm SIDM}}{m_{\rm DM}} \simeq \frac{\alpha^2_D \pi}{m^3_{\rm DM} v^4_{\rm DM}} \left({\rm log}\,R^2 -1\right) \, \lesssim \, 1 \, \frac{\rm cm^2}{\rm g} \, , 
\label{self_interaction bound}
\end{equation}
where $R = m_{\rm DM} v_{\rm DM} / m_V$ and $v_{\rm DM}$ is the DM relative velocity 
(taken to be $v_{\rm DM} \simeq 10^{-3}$ following \cite{Knapen:2017xzo}). 
Taking this upper-limit on $\alpha_D$ and the (combined) upper-bound on $\epsilon$ 
from \cite{Gaidau:2021vyr} (see also \cite{Caputo:2021eaa}) 
one can derive the DM self-interaction bound in the $\bar{\sigma}_e - m_{\rm DM}$ plane. 
Such a bound depends on the mass of dark photon mainly through 
the variation of the limit of $\epsilon$ with $m_V$. 

\medskip

The self-interaction bounds are presented in fig.~\ref{fig:constraints_mVultralight} by the 
red solid and dashed lines, that correspond to $m_V = 10^{-4}$ eV and 1 eV, respectively. 
While for $m_V = 1$ eV the self-interaction bound is much stronger than {\sc Damic-M}, 
for lower dark photon mass ($m_V = 10^{-4}$ eV) {\sc Damic-M} provides the leading constraints above 
$m_{\rm DM} \simeq 4$ MeV. 
As $m_{V} \rightarrow 0$, the self-interaction bound weakens and essentially evaporates, due to the increasing value of the limit on $\epsilon$; see, e.g., \cite{Caputo:2021eaa}. Thus, for the smallest mediator masses, 
DD experiments like {\sc Damic-M} are the most effective tool to probe the $\bar{\sigma}_e - m_{\rm DM}$ plane.

\section{Scalar portal: higgs-portal model}
\label{sec:model_scalar}

We also consider the scalar portal model for sub-GeV DM, 
where the interaction between DM (a Dirac fermion) 
and the Standard Model (SM) particles is mediated by a 
scalar that mixes with the SM higgs boson. 

The dark-sector Lagrangian for this model reads \cite{Baek_2012, Krnjaic_2016}: 
\begin{equation}
\mathcal{L_{D}}^S\, \supset \, \frac{1}{2}(\partial_{\mu}S\partial^{\mu}S-m_{S}^{2}S^{2})\, + \, \bar{\chi} (i \gamma^\mu \partial_{\mu}- g_{D}S- m_{\rm DM}) \chi \, + \, S\sin\theta\sum_{f}\frac{m_{f}}{v}\bar{f}f,
\label{lagrangian_higgs}
\end{equation}
with $\chi$ representing the Dirac DM particle, $S$ the scalar mediator, $\theta$ the mixing angle between the SM higgs and the mediator and $v =246$ GeV the higgs vacuum expectation value. We define, for a SM fermion $f$, $g_f = \sin\theta \, \frac{m_{f}}{v}$.
\newline
The reference DM-$e$ scattering cross-section is in this case~\cite{Xu:2024iny} 
(see Appendix~\ref{sec:sigma_e_higgsPortal}):
\begin{equation}
\bar{\sigma}_e = \frac{ g_e ^2 g_D^2 \mu^2_{{\rm DM}e}}
{\pi\left(\alpha ^2 m_e ^2+ m^2_{S}\right)^2} 
\label{eq:sigma_eHiggs}
\end{equation}
and the form factor

\[
F_{\rm DM}(q)=
\begin{cases}
\displaystyle 1 & \text{if } m_{S} \gg \alpha m_e \\
\displaystyle \left( \frac{\alpha \, m_e}{q} \right)^2  & \text{if } m_{S} \ll \alpha m_e
\end{cases}
\] 

\medskip

Analogously to what we did for the vector mediator model, below we discuss the cases with different values for the mass of the scalar mediator with respect to the DM mass. 
However, note that we skip the case with $m_S > m_{\rm DM}$, since 
in this case the limits from terrestrial searches (such as Higgs decays and rare meson decays) 
are much more constraining than both DD and ID bounds (see, e.g., \cite{Krnjaic_2016, Coogan_2021, Coogan:2021rez}).

\subsection{`Light' scalar mediator: $m_S < m_{\rm DM}$}
In this case, the $t$-channel process $\chi \bar{\chi} \to  S S$ is the dominant one and the velocity-averaged annihilation cross-section can be written as~\cite{Krnjaic_2016}:
\begin{equation}
\langle \sigma v \rangle = \frac{3g_{D}^{4}v_{\rm DM}^{2}}{128\pi \, m_{\rm DM}^{2}}
\end{equation}
where $v_{\rm DM}$ is the relative velocity of the DM particles. 
This ID cross-section is independent of the mixing angle $\theta$, 
contrary to $\bar{\sigma}_e$. Thus, the conversion of the ID bounds 
(on $\langle \sigma v \rangle$) to the $\bar{\sigma}_e - m_{\rm DM}$ plane 
requires an additional knowledge on the mixing angle which 
comes from terrestrial and astrophysical experiments, see \cite{Krnjaic_2016}.
In fact, one can make also a conservative conversion by choosing $\sin \theta = 1$. 
A smaller value of $\sin \theta$ (consistent with the experimental limits) 
would only strengthen the ID limits. 

\medskip

\begin{figure*}[!t]
\centering
\hspace{-2mm}
\includegraphics[width=0.50\textwidth]{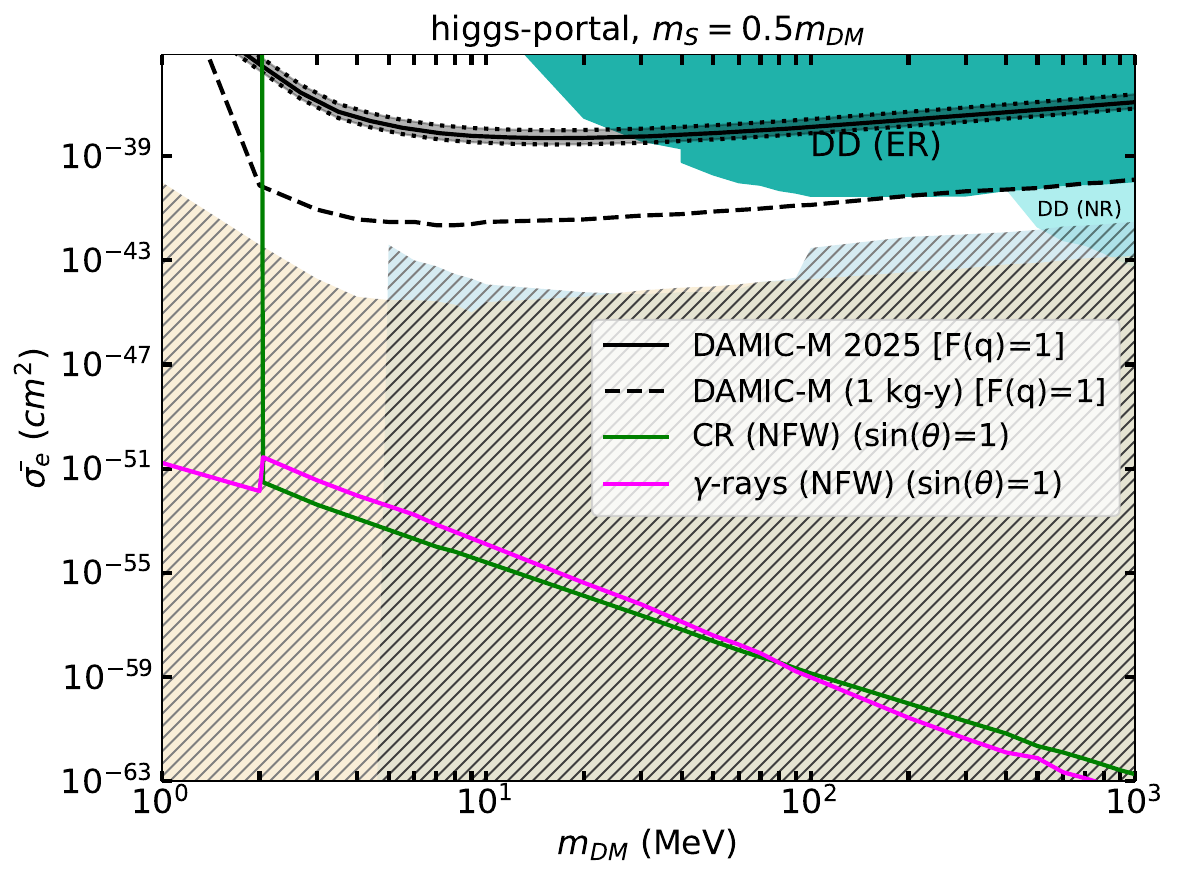}
\hspace{-2mm}
\includegraphics[width=0.50\textwidth]{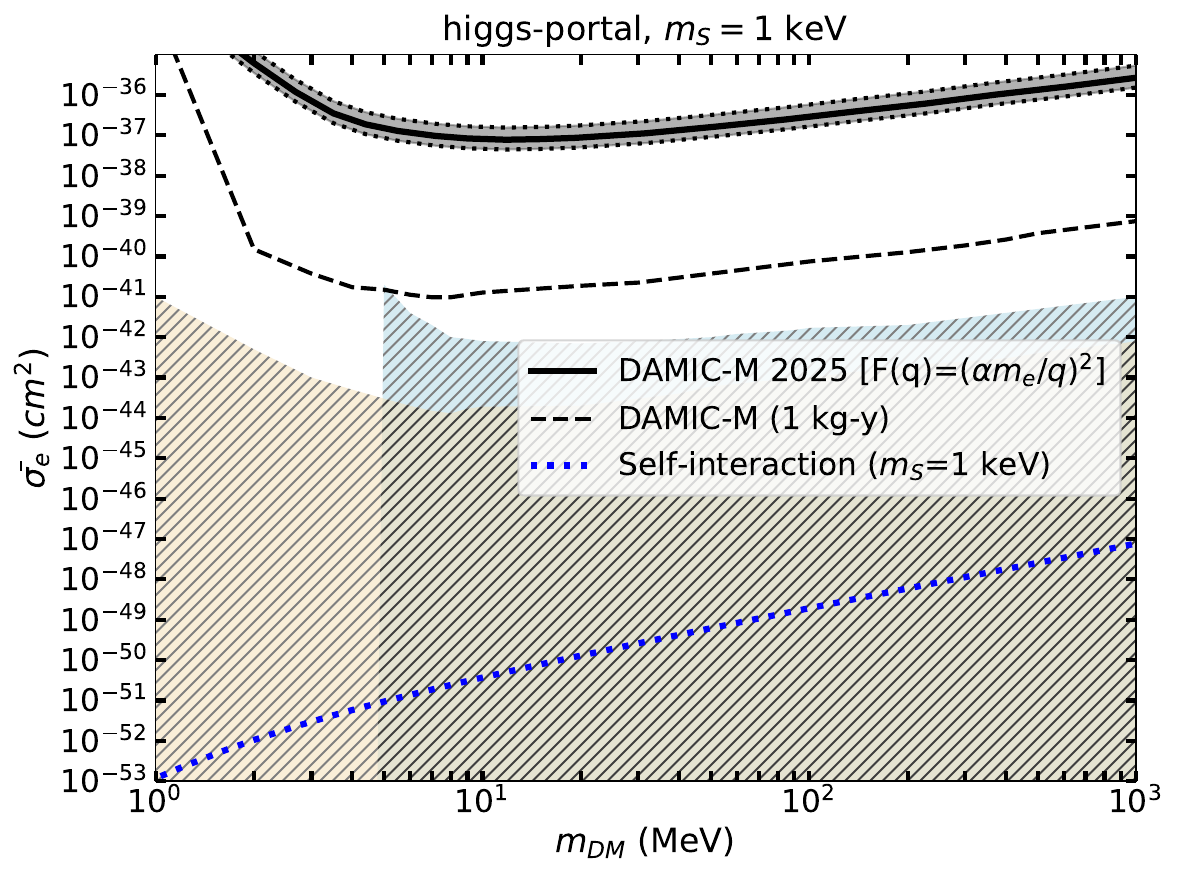}\\
\caption{\em Comparisons of ID and DD constraints for the {\bfseries higgs-portal model}, in the $\bar{\sigma}_e - m_{\rm DM}$ plane.
{\bfseries Left panel}: Limits for the case $m_S = m_{\rm DM}/2$. 
The {\sc Damic-M} bound from \cite{DAMIC-M:2025luv} 
(corresponding to $F(q) = 1$) is shown by the black line. 
The gray band corresponds to the variation of $\rho^{\rm DM}_\odot$ 
in the range [0.2 -- 0.7] $\rm GeV\,cm^{-3}$. The black dashed line shows the {\sc Damic-M} (1 kg-yr) 
projection from \cite{DAMIC-M:2022aks}. Note that the {\sc Damic-M} bound and projection are rescaled 
from $\rho^{\rm DM}_\odot =$ 0.3 to 0.4 ${\rm GeV.cm^{-3}}$. 
The ID bounds using CRs and $X$-rays/$\gamma$-rays are shown in green and magenta respectively 
(taken from \cite{Cirelli:2025rky}). The ID limits are plotted considering 
a maximal mixing ($\sin{\theta} = 1$). 
Areas excluded by nuclear and electron recoil DD experiments are shown in light and dark cyan respectively. 
{\bfseries Right panel:} Limits for the very light scalar portal case. 
The {\sc Damic-M} bound~\cite{DAMIC-M:2025luv} and projection~\cite{DAMIC-M:2022aks} 
correspond here to $F(q) = (\alpha m_e / q)^2$. The blue dotted line shows the DM self-interaction bounds for $m_S = 1$ keV.}
\label{fig:constraints_higgsportal}
\end{figure*}

For definiteness, we fix $m_S = m_{\rm DM} /2$ (as in \cite{Cirelli:2025rky}). 
The corresponding comparison between different experimental limits in the 
$\bar{\sigma}_e - m_{\rm DM}$ plane is provided in the left panel of 
fig.~\ref{fig:constraints_higgsportal}. The ID limits come from either 
the cosmic-ray observations (in green) or $X$-ray/$\gamma$-ray observations (in magenta). 
The DD bounds including the {\sc Damic-M} upper-limit and projection 
are the same as the ones shown in sec.~\ref{sec:light_DP}. 
The ID bounds (converted with $\sin \theta = 1$) are more constraining 
(at almost all DM masses) than that from the DD by several order of magnitudes. 
In fact, the regions probed by ID experiments are not accessible to DD experiments 
due to the neutrino floor~\cite{Cirelli:2024ssz} (shown by the hatched regions 
in fig.~\ref{fig:constraints_higgsportal}). More generally, 
this holds for both electron recoils (shown in dark cyan) and 
neutron recoils (in light cyan) experiments.

\subsection{`Very light' scalar portal: $m_S \ll 1$ keV}
Because of the same reasons of the dark photon case, here, too, 
the ID signal will be very much suppressed. 
$S$ being a scalar, there could be a signal from the process $\chi \bar{\chi} \rightarrow SS \rightarrow 2 \gamma (g) + 2 \gamma (g)$, but the very small decay rate of an ultralight $S$ into $\gamma$s or gluons makes this 
signal very suppressed too. 
On the other hand, like in the case of dark photon, there will be a strong 
limit from DM self-interactions, expressed as a bound on 
$\alpha_{\rm DM} = g^2_D / 4\pi$ via eq.~(\ref{self_interaction bound}). 
This limit can be converted and plotted in the $\bar{\sigma}_e - m_{\rm DM}$ plane, 
using additional knowledge on the mixing angle $\theta$, for example 
from~\cite{Hardy_2017}. 

We fix $m_S$ to 1 keV and show in the right panel of 
fig.~\ref{fig:constraints_higgsportal} that the self-interaction bounds 
are much stronger than the DD ones. This holds for all 
$m_S$ values in the range $\sim1$ eV -- 1 keV. 
For $m_S$ lower than $\sim$1 eV or higher than $\sim$1 keV, 
the self interaction limit becomes even stronger due to strengthening of 
the limit on $\sin{\theta}$~\cite{Hardy_2017}. Also, we use here the limit on 
$\sin{\theta}$ from \cite{Hardy_2017}; however using more stronger limits, 
e.g., those from \cite{Bottaro:2023gep, Fiorillo:2025zzx}, the self-interaction 
bound gets even stronger.

\section{Summary and conclusions}
\label{sec:conclusions} 

In this work we have combined direct detection (DD) and indirect detection (ID) searches 
of DM to analyze their complementarity in probing the 
parameter space of sub-GeV DM (with a mass in the range MeV -- GeV). 
This is done based on two representative realistic sub-GeV DM models, 
namely, the vector-portal and the scalar-portal models, where interactions 
between DM and SM are mediated by a vector and a scalar, respectively. 
For the vector-portal we focused on the dark photon which mixes 
kinematically with the photon, 
while for the scalar-portal we considered the higgs-portal scalar which mixes 
with the SM higgs.  

\medskip

Concerning DD, we focused in particular on the current set-up of the {\sc Damic-M} experiment, 
which uses a prototype detector with an exposure of $\sim$1.3 kg-day 
to constrain DM-electron scatterings, in the $\bar{\sigma}_e - m_{\rm DM}$ plane. 
Concerning the ID, we considered the fluxes of 
X-rays/$\gamma$-rays and cosmic-rays generated in the galaxy by the pair-annihilation of sub-GeV DM particles under the considered models and we derived updated constraints in the same $\bar{\sigma}_e - m_{\rm DM}$ plane, comparing 
with X-ray/$\gamma$-ray data from {\sc Integral}, {\sc Comptel}, {\sc Egret} and {\sc Fermi-Lat} and CR data from 
{\sc Voyager-1} and {\sc Ams-02}. 
For the photon signals we include, along with the prompt $\gamma$ emission, 
all possible secondary photons generated by the DM-induced $e^\pm$ via 
inverse compton, bremsstrahlung and in-flight annihilation processes, which help to improve the 
ID constraints at almost all DM masses (see, e.g., \cite{Cirelli:2025rky}). We include the possible 
variations in the ID constraints resulting from different astrophysical uncertainties 
(related to the variation of the galactic DM density and propagation of $e^\pm$ in the galaxy). 
In addition, we test the complementarity, in probing the sub-GeV DM parameter space, between the projected future results by {\sc Damic-M} (based on a larger exposure) and the upcoming space-based MeV telescope {\sc Cosi}. 
We also compare with existing bounds from other terrestrial, astrophysical and cosmological observations. 

\medskip

We found that the complementarity between 
{\sc Damic-M} and ID works best for the vector-portal model, especially 
with $m_V > m_{\rm DM}$ (we fix for definiteness $m_V = 3\,m_{\rm DM}$). 
In this case (see fig.~\ref{fig:constraints_RV3}), considering the different astrophysical uncertainties, 
the {\sc Damic-M} constraint reaches the same level as the ID bounds or even beats them for $m_{\rm DM}$ in the range $\sim3$--20 MeV. Outside this mass range, the ID bounds remain somewhat more constraining. 
In addition, the future {\sc Damic-M} will be able to provide a better sensitivity compared to {\sc Cosi} 
for $2\,{\rm MeV} \lesssim m_{\rm DM} \lesssim 80\,{\rm MeV}$, 
although outside this range {\sc Cosi} will provide a better probe. 
On the other hand, for the case $m_V < m_{\rm DM}$ 
(where we fix $m_V = m_{\rm DM}/3$ for definiteness), the present {\sc Damic-M} bound is much weaker 
than the ID bounds/projections, although the future {\sc Damic-M} can provide a comparatively better probe for a DM mass within the range 40--200 MeV (see fig.~\ref{fig:constraints_RV0.33}). 
In both of these cases, the ID observations often constrain the DM-$e$ scattering at levels below the $\nu$-floor, which the DD bounds cannot access.

We stress that, in both the above-mentioned cases for vector-portal DM, 
the present ID bounds and projections as well as the 
future {\sc Damic-M} projection probe uncharted parameter space, that is  
allowed by other DD, accelerator and astrophysical (SN1987A) searches, 
for almost the entire DM mass range MeV -- GeV. This stresses the importance of this kind of searches. 
Also, the present ID bounds already reach at the level of (or overcome) the CMB bound 
for different sub-GeV DM mass ranges. The upcoming {\sc Cosi} telescope as well as the future {\sc Damic-M} experiment will further improve on this. 

For the case with an ultralight $m_V$, the ID signals are very suppressed and thus irrelevant but  
DM self-interaction bounds become important. However, for $m_V \lesssim 10^{-4}$ eV, 
it is {\sc Damic-M} which provides the leading constraints in the $\bar{\sigma}_e - m_{\rm DM}$ plane 
for almost the entire MeV -- GeV DM mass range.  

\medskip

In the case of the higgs-portal model, the comparison between {\sc Damic-M} and ID is entirely different. Here the ID bounds based on existing data provide a much stronger constraint 
compared to {\sc Damic-M} for all DM masses, and exclude the full 
parameter space in which the DD experiments like {\sc Damic-M} are sensitive. 
In the case of an ultralight scalar mediator (where the ID bounds are very suppressed), 
the self-interaction bound entirely rules out the parameter space probed by {\sc Damic-M}.  

\bigskip

To summarize, we have found that, in a realistic sub-GeV DM model like the vector-portal scenario 
(with a mediator mass at/above the MeV scale), 
the DD bounds from {\sc Damic-M} and the ID bounds are comparable and complementary to each other 
(for some DM mass ranges) in probing the DM-$e$ scattering. 
Also, they are able to constrain new sub-GeV DM parameter space allowed by different other observations. 
The future updated version of {\sc Damic-M} and the upcoming MeV telescope {\sc Cosi} 
(as well as other similar MeV instruments like {\sc Amego}, {\sc e-Astrogam}, etc.) 
will be able to improve this situation even further.

\bigskip

\small
{\subsubsection*{Acknowledgments}
\footnotesize{
We thank Edoardo Vitagliano for 
pointing out the updates on supernova bounds and the 
self-interaction bounds concerning light scalar mediators. M.C.~and A.K.~acknowledge the hospitality of the Institut d'Astrophysique de Paris ({\sc Iap}) where part of this work was done. M.C.~also acknowledges the hospitality of the Theory Department at {\sc Cern}. 

\noindent Funding and research infrastructure acknowledgments: 
Research grant {\sl DaCoSMiG} from the {\sc 4eu+} Alliance (including Sorbonne Université); Institut Pascal and the {\sc P2i} axis of the Graduate School of Physics during the Paris-Saclay Astroparticle Symposium 2025. This project has also received financial support from the {\sc Cnrs} through the {\sc Miti} interdisciplinary programs.

\noindent H.L.~has received funding from the Agence Nationale de la Recherche ({\sc Anr} {\sl eDIM} project, {\sc Anr-24-Ce31-0381}). We thank the {\sc Damic-M}  collaboration for very fruitful discussions and for the use of their published limits and prospects. 
}}

\bigskip

\normalsize

\appendix

\section{Other vector portal models} 
\label{sec:other_vec_portal}

\begin{figure*}[!t]
\begin{minipage}{0.60\textwidth}
\hspace{-0.5cm}
\includegraphics[width=\textwidth]{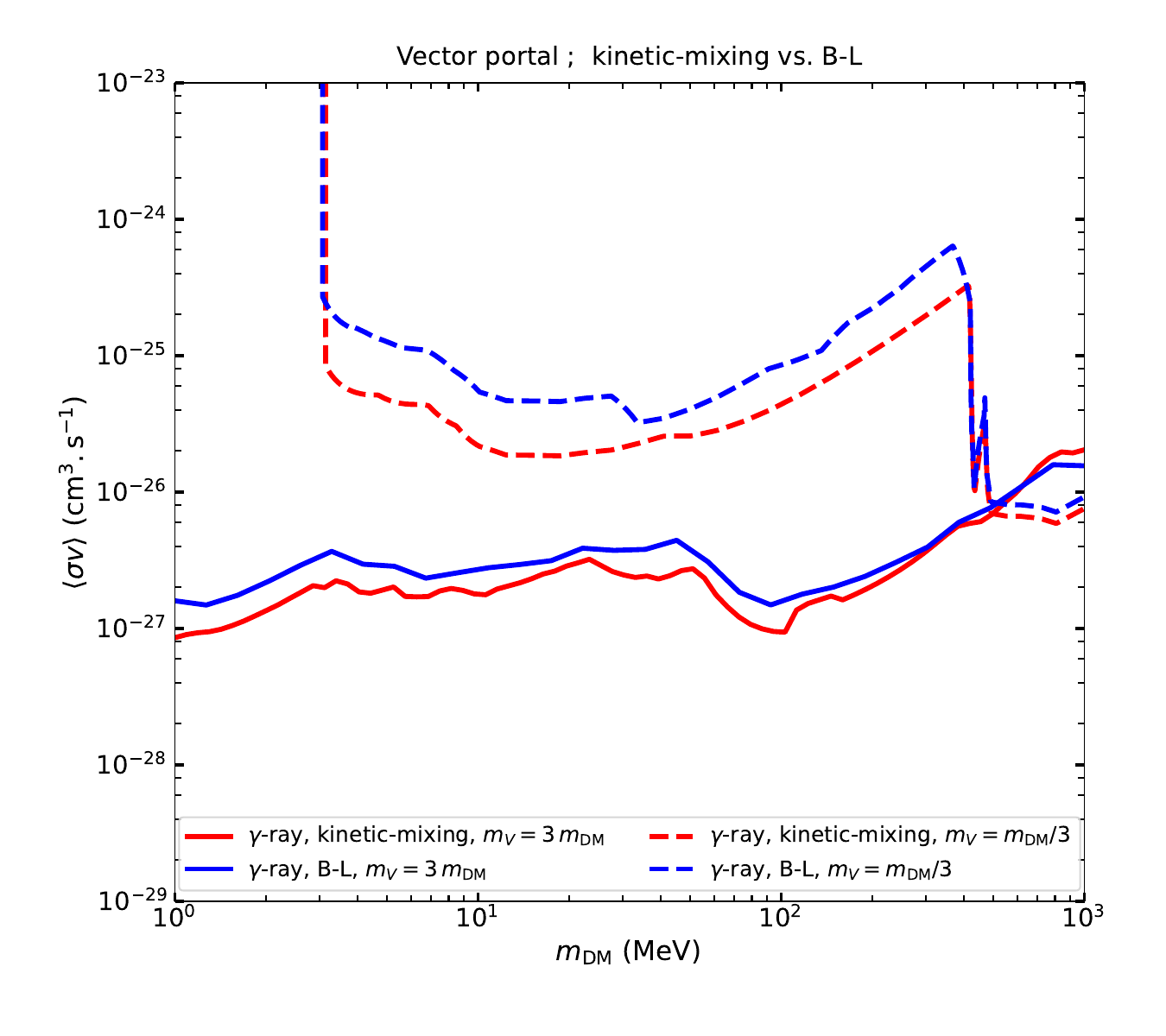}
\end{minipage}
\begin{minipage}{0.38\textwidth}
\vspace{-12mm}
\caption{\em {\bfseries Upper-limits} from $X$-ray/$\gamma$-ray observations are shown in the 
$\langle \sigma v \rangle - m_{\rm DM}$ plane {\bfseries for kinetic-mixing} (red) and {\bfseries the B-L models} (blue). The 
solid lines correspond to the heavy mediator case ($m_V = 3\,m_{\rm DM}$), while 
the dashed lines correspond to the light mediator case ($m_V = m_{\rm DM} / 3$). 
Here the benchmark 
NFW DM profile and the `conservative approach' (to obtain the ID limits using 
the astrophysical data) are considered.}
\label{fig:constraints_KM_BL}
\end{minipage}
\end{figure*}

In this section we briefly discuss the bounds on another vector portal model, namely the 
`B-L model' where the DM (taken to be a Dirac fermion) interacts with Standard Model particles 
via a light \(Z^{\prime }\) gauge boson that acts as a mediator 
(see e.g.~\cite{Coogan:2022cdd, Nath:2021uqb, Dutra:2025cwn} for details). 
This model has a similar structure of couplings between DM--mediator and mediator--SM-fermions 
like the dark photon model, apart from the additional coupling of the vector mediator 
with the neutral leptons. As a result, one can expect a similar type of bounds from ID observations and DD experiments. An example of this is illustrated in fig.~\ref{fig:constraints_KM_BL} where we compare the ID 
bounds on the dark photon model with that on the B-L model. As we can see the 
bounds on the two models are similar (within a factor of $\sim2$). 
Therefore, we keep our discussion in the main text focused on one vector portal scenario, 
the dark photon one. 

\section{Reference scattering cross section for the higgs-portal model}
\label{sec:sigma_e_higgsPortal}

This section provides a detailed calculation of the reference scattering cross section for the higgs-portal model. We consider the Lagrangian displayed in eq.~(\ref{lagrangian_higgs}).
For  the scattering process  $\chi(p_1)   e(k_1) \to\chi(p_2)  e(k2) $, the matrix element $\mathcal{M}$ reads:
\begin{equation}
  \mathcal{M} = \frac{-g_D g_e}{t - m_S^2}
  \bigl(\bar{u}(p_2)u(p_1)\bigr)\bigl(\bar{u}(k_2)u(k_1)\bigr)
\end{equation}
with  $t = (p_1 - p_2)^2 = (k_2 - k_1)^2$ the squared transferred momentum.
\newline
The squared matrix element $|\mathcal{M}|^2$  averaged on spins is therefore:
\begin{equation}
  \overline{|\mathcal{M}|^2} =\frac{1}{4}\sum_{\text{spins}}|\mathcal{M}|^2=\frac{g_D^2 g_e^2}{4(t-m_S^2)^2}
\left(\sum_{\text{spins}}|\bar{u}(p_2)u(p_1)|^2\right)
 \left(\sum_{\text{spins}}|\bar{u}(k_2)u(k_1)|^2\right)
\end{equation}
Using $\sum_s u^s(p)\bar{u}^s(p) = \slashed{p} + m$, we get 
\begin{equation}
  \frac{1}{2}\sum_{\text{spins}}|\bar{u}(p_2)u(p_1)|^2
  = \frac{1}{2}\,\mathrm{Tr}\bigl[(\slashed{p}_2+m_{\rm DM})(\slashed{p}_1+m_{\rm DM})\bigr]
\end{equation}
and by using  ${\rm Tr}({\gamma}^{\mu})=0$ and ${\rm Tr}({\gamma}^{\mu} {\gamma}^{\nu})=4{\eta}^{\mu \nu}$, we obtain
\begin{equation}
  \frac{1}{2}\sum_{\text{spins}}|\bar{u}(p_2)u(p_1)|^2 = 2(p_1\cdot p_2 + m_{\rm DM}^2).
\end{equation}
In the same way, 
\begin{equation}
  \frac{1}{2}\sum_{\text{spins}}|\bar{u}(k_2)u(k_1)|^2 = 2(k_1\cdot k_2 + m_e^2).
\end{equation}

The typical velocities of DM particles in the halo are of the order $v\sim 10^{-3}$. In the non-relativistic regime ($|\vec{p}|\ll m$), we  arrive at
\begin{equation}
    \overline{|\mathcal{M}|^2}
  = \frac{16 g_D^2 g_e^2 \,m_{\rm DM}^2 m_e^2}
         {(q^2 + m_S^2)^2}
\end{equation}
where $\vec{q} = \vec{p}_1 - \vec{p}_2$.
\newline 
Using the definition of  $\bar{\sigma}_e$ (see \cite{Essig_2012}), one finally has: 
\begin{equation}
    \bar{\sigma}_e
  =
  \frac{
  \mu_{\rm {DM} e}^2 \,
  g_D^2
  g_e^2
  }
  {
  \pi\,
  \left(\alpha^2m_e^2+m_S^2\right)^2
  }
\end{equation}
\begin{equation}
    |F_{\rm DM} (q)|^2 =\frac{\left(\alpha^2m_e^2+m_S^2\right)^2}{\left(q^2+m_S^2\right)^2}
\end{equation}
where $ \alpha m_e$ is the typical transferred momentum in  {\sc Damic-M}-like experiments.

\clearpage

\bibliographystyle{JHEP}
\bibliography{bibliography.bib}

\end{document}